\documentclass[12pt]{msml2020} %

\usepackage{amsmath,amsfonts,bm}

\def\eqref#1{equation~\ref{#1}}

\def\1{\bm{1}}

\DeclareMathAlphabet{\mathsfit}{\encodingdefault}{\sfdefault}{m}{sl}
\SetMathAlphabet{\mathsfit}{bold}{\encodingdefault}{\sfdefault}{bx}{n}

\DeclareMathOperator*{\argmax}{arg\,max}

\usepackage{subcaption}
\usepackage{hyperref}
\usepackage{url}
\usepackage{placeins}
\usepackage{wrapfig}
\usepackage{caption}
\usepackage{nicefrac}
\usepackage{microtype}
\usepackage{graphicx}
\usepackage{float}
\usepackage{booktabs} %
\usepackage{algorithm}
\usepackage{algorithmic}
\usepackage{hyperref}
\usepackage{physics}

\definecolor{mydarkblue}{rgb}{0,0.08,0.45}

\hypersetup{
    unicode=false,          %
    pdftoolbar=true,        %
    pdfmenubar=true,        %
    pdffitwindow=false,     %
    pdfstartview={FitH},    %
    pdftitle={My title},    %
    pdfauthor={Author},     %
    pdfsubject={Subject},   %
    pdfcreator={Creator},   %
    pdfproducer={Producer}, %
    pdfkeywords={keyword1, key2, key3}, %
    pdfnewwindow=true,      %
    colorlinks=true,       %
    linkcolor=mydarkblue,          %
    citecolor=mydarkblue,        %
    filecolor=mydarkblue,      %
    urlcolor=mydarkblue           %
}

\newcommand{\params}{{\bm\theta}}
\numberwithin{equation}{section}

\newcommand{\aref}[1]{\hyperref[#1]{Appendix~\ref*{#1}}}

\def\equationautorefname~#1\null{Eq.~(#1)\null}

\usepackage{subcaption}

\newcommand{\I}{\mathrm{i}}

\usepackage[normalem]{ulem}

\def \agentname {PG-QAOA}

\title[PG-QAOA]{\hspace{-1mm}Policy Gradient based Quantum Approximate Optimization Algorithm}
\usepackage{times}

\msmlauthor{%
 \Name{Jiahao Yao} \\
 \addr Department of Mathematics, University of California, Berkeley,  CA 94720, USA
 \AND
 \Name{Marin Bukov} \\
 \addr Department of Physics, University of California, Berkeley, CA 94720, USA
 \AND
 \Name{Lin Lin} \\
 \addr Department of Mathematics, University of California, Berkeley,  CA 94720, USA;\\ Computational Research Division, Lawrence Berkeley National Laboratory, Berkeley, CA 94720, USA
}

\begin{document}

\maketitle

\begin{abstract}%
The quantum approximate optimization algorithm (QAOA), as a hybrid quantum/classical algorithm, has received much interest recently. QAOA can also be viewed as a variational ansatz for quantum control. 
However, its direct application to emergent quantum technology encounters additional physical constraints: (i) the states of the quantum system are not observable; (ii) obtaining the derivatives of the objective function can be computationally expensive or even inaccessible in experiments, and (iii) the values of the objective function may be sensitive to various sources of uncertainty, as is the case for noisy intermediate-scale quantum (NISQ) devices.
Taking such constraints into account, we show that policy-gradient-based reinforcement learning (RL) algorithms are well suited for optimizing the variational parameters of QAOA in a noise-robust fashion, opening up the way for developing RL techniques for continuous quantum control.
This is advantageous to help mitigate and monitor the potentially unknown sources of errors in modern quantum simulators. 
We analyze the performance of the algorithm for quantum state transfer problems in single- and multi-qubit systems, subject to various sources of noise such as error terms in the Hamiltonian, or quantum uncertainty in the measurement process. We show that, in noisy setups, it is capable of outperforming state-of-the-art existing optimization algorithms. 
\end{abstract}

\begin{keywords}%
  Quantum approximate optimization algorithm, Policy gradient, Reinforcement learning, Robust optimization, Quantum computing, Quantum control%
\end{keywords}

\section{Introduction}

Noisy intermediate scale quantum (NISQ) devices are becoming increasingly accessible. However, their performance can be severely restricted due to decoherence effects. This leads to noises in all components of the quantum computer, including initial state preparation, unitary evolution, and measurement/qubit readout. Thanks to the feasibility of being implemented and tested on near term devices, hybrid quantum-classical algorithms, and in particular quantum variational algorithms (QVA), have received significant amount of attention recently. Examples of QVA include the Variational Quantum Eigensolver \citep{peruzzo2014variational}, the Quantum Approximate Optimization Algorithm (QAOA)  \citep{farhi2014quantum},  Quantum Variational Autoencoders \citep{romero2017quantum}, etc. The common feature of these algorithms is that the final wavefunction can be prepared by applying a unitary evolution operator, parametrized using a relatively small number of parameters, to an initial wavefunction. The parameters can then be variationally optimized to maximize a given objective function, measured on the quantum device.

In this study, we mainly focus on the Quantum Approximate Optimization Algorithm (QAOA) \citep{farhi2014quantum}, which is a particularly simple algorithm that alternates between two different unitary time evolution operators of the form $e^{-\I H_{0}t}$, $e^{-\I H_{1}t}$ ($t\in \mathbb{R}$). This is also dubbed as the quantum alternating operator ansatz~\citep{hadfield2019quantum}. Both share the same acronym QAOA. We use the term QAOA in a broader sense than that in the original paper by Farhi et al. The algorithm proposed here can be potentially used for a larger class of variational quantum circuits (VQC, which includes the variational quantum eigensolver, VQE as a special case). However, it would be practically difficult to assess the efficiency and robustness of the method for general VQCs. Therefore for concreteness, in this work we specifically confine our study to QAOA.

QAOA has been studied in the context of a number of discrete~\citep{farhi2014quantum,lloyd2018quantum,hadfield2018quantum} and continuous ~\citep{verdon2019quantum} optimization problems. QAOA has also been demonstrated to be universal under certain circumstances \citep{lloyd1995almost,lloyd2018quantum,morales2019universality}, in the sense that any element in a unitary group can be well approximated by a properly parameterized QAOA. This is highly nontrivial and is a unique quantum feature, since QAOA only has access to  unitary operators generated by two specific Hamiltonians $H_0,H_1$. However, the control energy landscape of QAOA is known to be highly complex \citep{streif2019training,niu2019optimizing}, and optimization in it can therefore be challenging. 
For a one-level system, the QAOA optimization landscape in a channel decoding problem can already be quite complex~\citep{matsumine2019channel}.
For random parameterized quantum circuits (RPQCs), the average value of the gradient of the objective function has been reported to be almost zero~\citep{mcclean2018barren}. Such vanishing gradients in large plateaus pose challenge to optimization algorithms.  
If the landscape has exponentially many local minima, there is exponentially small probability of reaching the global minimum~\citep{day2019glassy}.

QAOA can be naturally related to quantum control, and thus also to reinforcement learning problems. This inspires studies from various angles, such as the Krotov method~\citep{tannor1992control}, Pontryagin’s maximum principle ~\citep{yang2017optimizing} and Bayesian optimization~\citep{Sauvage2019}, sequential minimal optimization~\citep{nakanishi2019sequential}, tabular reinforcement learning methods ~\citep{chen2013fidelity,bukov2018quantum}, functional-approximation-based (deep) Q-learning methods ~\citep{bukov2018reinforcement, sordal2019deep, an2019deep, zhang2019reinforcement}, policy gradient methods ~\citep{fosel2018reinforcement,august2018taking, chen2019manipulation, niu2019universal, porotti2019coherent, wauters2020reinforcement}, differential programming~\citep{Schafer2020} and methods inspired by the success of AlphaZero~\citep{dalgaard2019global}. 
Most studies focus on the noise-free scenarios, applicable to fault-tolerant quantum devices. In order to mitigate the errors on near-term devices, robust optimization based on sequential convex programming (SCP) has been recently studied \citep{kosut2013robust,dong2019robust}, which assumes that both the source and the range of magnitude of the error are known, but its exact magnitude. In such a case, the authors have found that robust optimization can significantly improve the accuracy of the variational solution. 

Nonetheless, techniques such as SCP require access to information of the first as well as second order derivatives of the objective function, which can themselves be noisy and difficult to obtain on quantum devices. The objective function should also be at least continuous with respect to the error, a requirement which is not satisfied in the case of quantum uncertainty in the final measurement process (e.g. in the form of a bit flip or a phase flip). It is thus naturally desirable to only use function evaluations to perform robust optimization, while keeping the result resilient to unknown and generic types of errors. 

In this paper, we demonstrate that reinforcement learning (RL) may be used to tackle all challenges above in optimizing the parameters of QAOA, and more generally QVA. Instead of directly optimizing the variational parameters themselves, we may assign a \textit{probability distribution} to the parameter set, and perform optimization with respect to the parameters of the probability distribution, denoted $\params$. The modified objective function (called the expected total reward function) can then be continuous with respect to $\params$, even if the original objective function is not. The optimization procedure only requires a (possibly large) number of function evaluations, but no information about the first or second order derivatives. We show that a simple policy gradient method only introduces a small number of additional parameters in the optimization procedure, and can be used to optimize the parameters in QAOA. Since each step of the optimization only involves a small batch of samples, the optimization procedure can also be resilient to various sources of noise.

This paper is organized as follows. Section \ref{sec:QAOA} provides a brief introduction of QAOA, its connection to quantum control, and the noise models. Section \ref{sec:pgqaoa} introduces the policy gradient based QAOA (PG-QAOA), in the context of noise-free and noisy optimization. After introducing the test systems in Section \ref{sect:models},  we present in Section \ref{sect:errmitigate} numerical results of PG-QAOA for  single-qubit and multi-qubit examples under different noise models. Section \ref{sect:sum} concludes and discusses the further work.
Additional numerical results are presented in the Appendices.

\section{Preliminaries}\label{sec:QAOA}
\subsection{QAOA and Quantum Control}

Consider the Hilbert space $\mathcal{H}=\mathbb{C}^{2^N}$, with $N$ the number of qubits in the quantum system. 
Starting from an initial quantum state $\ket{\psi_i}\in \mathcal{H}$, in QAOA we apply two alternating unitary evolution operators \citep{farhi2014quantum}:
\begin{equation}
\ket{\psi}=U(\{\alpha_i, \beta_i\}_{i=1}^p) \ket{\psi_i}= e^{-\I H_1 \beta_p} e^{-\I H_0 \alpha_p} \cdots e^{-\I H_1 \beta_1} e^{-\I H_0 \alpha_1}\ket{\psi_i}.
\label{eqn:qaoawf}
\end{equation} 
The unitary evolution is generated by the time-independent Hamiltonian operators $H_0$ and $H_1$, each applied for a duration $\alpha_i\geq0$ and $\beta_i\geq0$, respectively  ($i=1,2,\cdots, p$); we refer to $p$ as the total depth. In QAOA, we have to adjust the parameters to optimize an objective function $F(\ket{\psi})=F(\{\alpha_i, \beta_i\}_{i=1}^p)$, e.g.~minimizing the energy \citep{ho2019efficient} or maximizing the fidelity of being in some target state\footnote{The latter problem is often referred to as a state transfer problem. This is mainly for simplicity and serves as a proof of principle for the effectiveness of the policy gradient based method. }. 
In the latter case, for a target wavefunction denoted by $\ket{\psi_\ast}$, the optimization problem becomes
\begin{eqnarray}
    \{\alpha_i^\ast, \beta_i^\ast\}_{i=1}^p &=& \argmax_{\{\alpha_i, \beta_i\}_{i=1}^p} F(\{\alpha_i, \beta_i\}_{i=1}^p),\\
    F(\{\alpha_i, \beta_i\}_{i=1}^p) &=& \left\vert \mel{\psi_\ast}{U(\{\alpha_i, \beta_i\}_{i=1}^p)}{ \psi_i} \right\vert^2.    
\label{eqn:qaoa}
\end{eqnarray}

The problem of finding the optimal parameters in QAOA can be reinterpreted as the following bilinear quantum optimal control problem
\begin{equation}
\I \partial_t \ket{\psi(t)}=H(t)\ket{\psi(t)}, \quad \ket{\psi(0)}=\ket{\psi_i},
\label{eqn:optimalcontrol}
\end{equation}
where $H(t)=H_0+u(t)(H_1-H_0),u(t)\in\{0,1\}$. In particular, when $u(t)$ is chosen to be the following piecewise constant function
\begin{equation}
 u(t)=\begin{cases}
 0, & t\in \big[\sum_{k=1}^{i-1}(\alpha_k+\beta_k),\sum_{k=1}^{i-1}(\alpha_k+\beta_k)+\alpha_i\big),\\
 1, & t\in \big[\sum_{k=1}^{i-1}(\alpha_k+\beta_k)+\alpha_i,\sum_{k=1}^{i}(\alpha_k+\beta_k)\big),\\
 \end{cases} \quad i=1,\ldots,p,
\end{equation}
we recover the QAOA wavefunction (\ref{eqn:qaoawf}). This is a special type of quantum control problem called the bang-bang quantum control. For a protocol of the durations  $\{\alpha_i, \beta_i \}_{i=1}^p$, the total duration is defined as 
\begin{equation}
    T\left(  \{\alpha_i, \beta_i \}_{i=1}^p \right)= \sum_{i=1}^p \left( \alpha_i + \beta_i \right).
    \label{eqn:duration}
\end{equation}

\subsection{Noisy Objective Functions}\label{subsec:QAOA_noise}

Practical QAOA calculations can be prone to noises. For instance, the Hamiltonian may take the form $H(\delta) = \bar{H}+\delta \tilde{H}$, where $\bar{H}$ is the  Hamiltonian in the absence of noise,  $\tilde{H}$ is the Hamiltonian modelling the noise source, with $\delta$ the magnitude of the noise. We assume that only the range/magnitude of $\delta$ is known \textit{a priori} and is denoted by $\Delta$, and the precise value of $\delta$ is not known. This setup will be referred to as the  \textit{Hamiltonian noise}. The explicit form of the Hamiltonian noise will be discussed later in Section~\ref{set:multi-qubits}.
This noisy optimization problem can be solved as a max-min problem: 

\begin{equation}
\label{eq:control}
\begin{array}{cc}
\max\limits_{\{\alpha_i, \beta_i\}_{i=1}^p}  &  \min\limits_{\delta\in \Delta} {F}(\{\alpha_i, \beta_i\}_{i=1}^p, \delta),
\end{array}
\end{equation}
where
\begin{equation}
{F}(\{\alpha_i, \beta_i\}_{i=1}^p, \delta) = | \mel{\psi_i}{U(\{\alpha_i, \beta_i\}_{i=1}^p, \delta)}{\psi_\ast}|^2
\end{equation}
is the fidelity for the given noise strength and control duration. %

Noise may naturally also occur due to imperfect  measurement operations. For instance, the final fidelity may only be measurable up to an additive Gaussian noise, i.e. 
\begin{equation}
    F_\sigma (\{\alpha_i, \beta_i\}_{i=1}^p)= \operatorname{clip}(F(\{\alpha_i, \beta_i\}_{i=1}^p) + \epsilon, 0, 1), 
    \label{eqn:gsfid}
\end{equation}
where $\epsilon \sim \mathcal N (0, \sigma^2)$. Here, the $\operatorname{clip}$\footnote{This is just one way to enforce the fidelity to be between 0 and 1. We admit that the clipping procedure is an artifact of the implementation; it is not necessary and does not constitute an essential feature of the PG-QAOA algorithm.} function guarantees the noisy fidelity is still bounded between 0 and 1. This will be referred to as \textit{Gaussian noise}. It mimics the case when experimentalists do lots of measurements and average the result in the end to get an estimate of the observable. By the central limit theorem, as the sample size becomes sufficiently large, the statistics of the measurement data points is well approximated by a Gaussian distribution. As a result, we use the Gaussian noise to describe the uncertainty leading to a noise in the reward signal.

Furthermore, quantum measurements produce an intrinsic source of uncertainty due to the probabilistic nature of quantum mechanics. Assuming the target state $|\psi_\ast\rangle$ is an eigenstate of some measurable operator $O$ with eigenvalue $o_\ast$, i.e.~$O|\psi_\ast\rangle = o_\ast|\psi_\ast\rangle$, a quantum measurement $\bra{\psi}O\ket{\psi}$ produces the eigenvalue $o_\ast$ with probability $F(\{\alpha_i, \beta_i\}_{i=1}^p)$. Using this, we can define the following discrete cost function:
\begin{equation}
F_Q (\{\alpha_i, \beta_i\}_{i=1}^p)=\left\{\begin{array}{ll}{1} & {\text { with probability } F (\{\alpha_i, \beta_i\}_{i=1}^p)} \\ {0} & {\text { with probability } 1-F (\{\alpha_i, \beta_i\}_{i=1}^p)}\end{array}\right.
\label{eqn:qfid}
\end{equation}
Assuming the same state $\ket{\psi}$ of the system is prepared anew in a series of experiments, a measurement in repeated experiments will produce a discrete set of ones and zeros, whose mean value converges to the true fidelity $F (\{\alpha_i, \beta_i\}_{i=1}^p)$ in the limit of large number of quantum measurements. 
This setting was considered in  \citep{bukov2018quantum}, and will be referred to as the \textit{quantum measurement noise}. We mention in passing that, in systems with large Hilbert spaces, such as multi-qubit systems, it is in fact more appropriate to optimize the expectation value of some observable, instead of the fidelity.

\section{Policy gradient based QAOA (PG-QAOA)}\label{sec:pgqaoa}

Being a variational ansatz, QAOA does not specify the optimization procedure to determine the variational parameters. In this paper, we demonstrate that policy gradient, which is a widely used algorithm in reinforcement learning, can be particularly robust to various sources of uncertainty in the physical system.
In order to tackle the robust optimization of QAOA for general noise models,  reinforcement learning algorithms provide a useful perspective. 

We first reformulate the original problem as a  probability-based optimization problem.  The original optimization parameters are drawn from a probability distribution, described by some variational parameters $\params$. Optimization is then performed over the variational parameters. Such techniques are used in neural evolution strategies (NES) \citep{wierstra2008natural} and model-based optimization (MBO) \citep{brookes2019view}. It is also shown very recently by \citep{zhao2020natural} that NES can be efficiently applied to solve combinatorial optimization problems in the quantum classical approach.  If the solution of the optimization problem is unique, we should expect that the probability distribution of each parameter will converge to a Dirac-$\delta$ distribution. This probability-based approach also has the advantage of being resilient to perturbations and noise. As will be shown later, the width of the probability distribution after optimization can also be used to qualitatively monitor the magnitude of the unknown noise. A common example of a probability-based optimization algorithm in reinforcement learning is the policy gradient algorithm, where the goal is to find the optimal  policy $\pi_\params$ to perform a given task \citep{williams1992simple, sutton2018reinforcement}.
An additional advantage of probability-based optimization is that it can be used to handle continuous and discrete variables in a unified fashion. Thus, the ideas we put forward below can be used to open up the way to applying RL techniques to continuous quantum control. In the context of QAOA, the durations can be treated as continuous variables without the error due to time discretization.

Let us begin by casting the QAOA control problem \autoref{eqn:duration} in a reinforcement learning framework. We consider finite-horizon episodic learning task, with $p$ steps per episode, naturally defined by the discrete character of the QAOA ansatz \autoref{eqn:qaoa}.

The natural choice for the RL state space is the Hilbert space $\mathcal{H}$. However, there are a number of problems associated with this choice: (i) the wave function $\ket{\psi}$ is a mathematical concept, which cannot be directly measured in experiments (for instance, there is an arbitrary phase factor that cannot be directly measured). (ii) in quantum mechanics, every measurement would directly lead to the collapse of the wavefunction. (iii) in many-body quantum systems of $N$ particles, $\dim \mathcal{H}\sim \exp(N)$ is exponentially large which raises questions about the scalability of the algorithm to a large number of qubits. Indeed, reading out all entries of the quantum wavefunction requires full quantum tomography~\citep{torlai2018neural}, which scales exponentially with the number of qubits $N$. This comes in stark contrast with recent applications of RL to certain optimal quantum control problems e.g.~\citep{niu2019universal,dalgaard2019global}, in which the quantum wavefunction for small Hilbert spaces is indeed accessible on a classical computer.

In our setting, since the dynamics is governed by the Schrodinger equation and initial state is also given, the quantum state at an intermediate time $\ket{\psi(t)}$ can be in principle determined from the sequence of actions taken at each time interval. Therefore, the sequence of all actions taken before a given episode step can be treated effectively as the RL state, and we work with this definition here. We mention in passing that this choice is not unique: in practice, reinforcement learning based methods often incorporate some form of embedding of the quantum state as their state. Notable examples include tabular Q-Learning \citep{bukov2018reinforcement}, Q-Learning network \citep{sordal2019deep, an2019deep}, LSTM based memory proximal policy optimization \citep{august2018taking,fosel2018reinforcement}.

At every step $j$ in the episode, our RL agent is required to choose two actions out of a continuous interval $[0,\infty)$ independently, representing the values of the durations $\alpha_j,\beta_j$. Hence, the action space is $\mathcal{A}=[0,\infty)$. Actions are selected using the parameterized policy $\pi_{\params}$. Since we use the fidelity as the objective function, the reward space is $\mathcal R = [0, 1]$.

In this work, we use the simplest ansatz, i.e. independent Gaussian distributions to parameterize the policy over the control durations $\{\alpha_i, \beta_i\}_{i=1}^p$ in QAOA. Since a Gaussian is uniquely determined by its mean $\mu$ and standard deviation (std) $\sigma$, we have a total of $2p$ independent variational parameters ${\bf\params} = \{\mu_{\alpha_i}, \sigma_{\alpha_i}, \mu_{\beta_i}, \sigma_{\beta_i}\}_{i=1}^p$. The total number of parameters is $4p$ (in particular, it does not directly scale with the number of qubits $N$). The probability density of all the parameters $\pi_{\params}(\{\alpha_i, \beta_i\}_{i=1}^p)$ is the product of all the marginal distributions:
\begin{equation}
    \pi_{\params}(\{\alpha_i, \beta_i\}_{i=1}^p) = \prod_{i=1}^p \pi(\alpha_i ;\mu_{\alpha_i}, \sigma_{\alpha_i})\cdot \pi(\beta_i ;\mu_{\beta_i}, \sigma_{\beta_i}),
    \label{eqn:prob}
\end{equation}
where $\pi(x; \mu, \sigma)$ is the probability density for the Gaussian distribution $\mathcal N(\mu, \sigma)$, 
\begin{equation}
\pi(x; \mu, \sigma) = \frac{1}{\sqrt{2\pi \sigma^2}}e^{- \frac{(x-\mu)^2 }{2\sigma^2}}.
\label{eqn:Gaussian}
\end{equation}
Note that with such a choice, $x$ may become negative, which lies outside the action space $\mathcal{A}$. We can enforce the constraint using a truncated Gaussian distribution (after proper normalization) or a log-normal distribution. In practice we observe that with proper initialization, the positivity condition is observed to be automatically satisfied by the minimizer even with the simple choice in Eq. (\ref{eqn:Gaussian}).

\begin{figure}[ht]
    \centering
    \includegraphics[width=0.8\textwidth]{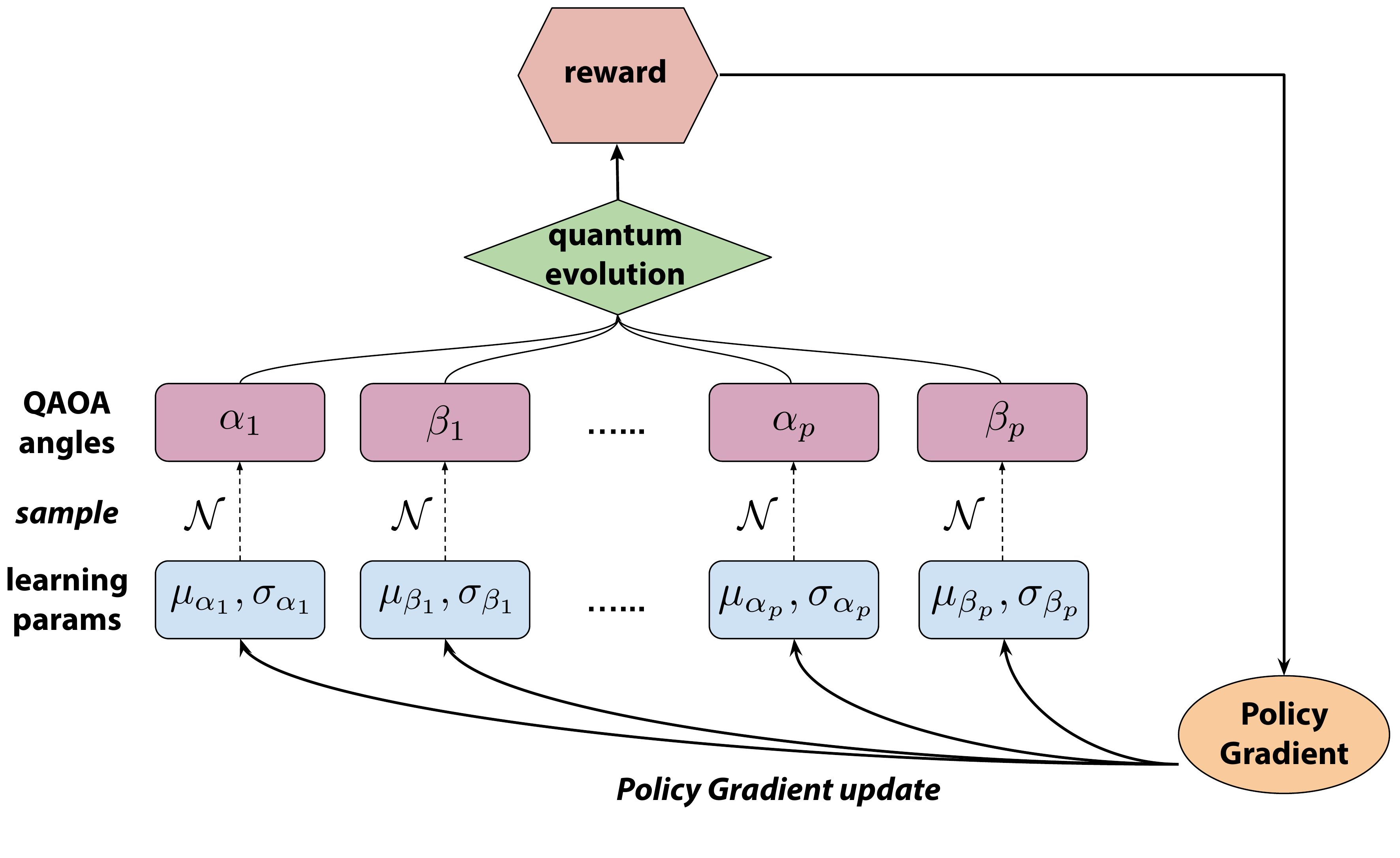}
    \caption{\small Schematic diagram for \agentname{}. 
    The algorithm samples a batch of QAOA time-durations (angles) from the current policy, aggregates the resulting fidelities/rewards from a quantum `blackbox', and applies the policy gradient algorithm to update the learning parameters to improve the policy. }
	\label{fig:pgqaoa-arch-diag}
	\vspace{-2mm}
 \end{figure}
The QAOA objective function~(\ref{eqn:qaoa}) for the probability-based ansatz~(\ref{eqn:prob}) introduced above, now takes the form:

\begin{equation}
    \{\mu_{\alpha_i}^*, \sigma_{\alpha_i}^*, \mu_{\beta_i}^*, \sigma_{\beta_i}^*\}_{i=1}^p = 
    \argmax_{\{\mu_{\alpha_i}, \sigma_{\alpha_i}, \mu_{\beta_i}, \sigma_{\beta_i}\}_{i=1}^p} 
    \left(
    \mathop{ \mathbb E}_{\substack{\alpha_i\sim \mathcal N(\mu_{\alpha_i}, \sigma_{\alpha_i}) \\ \beta_i\sim \mathcal N(\mu_{\beta_i}, \sigma_{\beta_i}) }}
    \left[ F(\{\alpha_i, \beta_i\}_{i=1}^p) \right]
    \right)
    = \argmax_{\bf\params} J({\bf\params}). 
    \label{eqn:probopt}
\end{equation}
Here $J(\params)$ is called the expected reward function. 
In this form, the objective function $J({\bf\params})$ can be optimized using the REINFORCE algorithm for policy gradient \citep{williams1992simple}:
\begin{equation}
\nabla_{\params}J(\params) = 
\mathop{ \mathbb E}_{\substack{\alpha_i\sim \mathcal N(\mu_{\alpha_i}, \sigma_{\alpha_i}) \\ \beta_i\sim \mathcal N(\mu_{\beta_i}, \sigma_{\beta_i}) }}
\left[\nabla_{\params}\log \pi_{\params}(\{\alpha_i, \beta_i\}_{i=1}^p)
\cdot
F(\{\alpha_i, \beta_i\}_{i=1}^p)
\right],
\label{eqn:pg}
\end{equation}

In particular, the gradient can be evaluated without information about the first order derivative of the objective function $F$. In practice, we use a Monte Carlo approximation to evaluate this gradient, as shown in \autoref{alg:probopt}.  
In order to reduce the variance of the gradient, usually a baseline 
is subtracted from the fidelity~\citep{greensmith2004variance}, i.e. replacing $F(\{\alpha_i, \beta_i\}_{i=1}^p)$ with $F(\{\alpha_i, \beta_i\}_{i=1}^p) - \bar F$ in  \autoref{eqn:pg}; it is easy to compute the average fidelity over the MC sample (i.e.~the batch) and we use that as the baseline. The resulting algorithm will be referred to as the policy gradient based QAOA (PG-QAOA). 
\begin{algorithm}[H]
  \caption{Policy gradient based QAOA}
  \label{alg:probopt}
  \begin{algorithmic}[1]
\REQUIRE Initial guess for the mean and std $\mu_{\alpha_i}^0, \sigma_{\alpha_i}^0, \mu_{\beta_i}^0, \sigma_{\beta_i}^0, i =1,2,\cdots, p;$ \\ \quad \; batch size $M$, learning rate $\tau_t$, total number of iterations $N_{\text{iter}}$.\\ 
  \STATE Initialize the mean and std with the initial guess 
       \vspace{-0.6em}
  $$(\mu_{\alpha_i}, \sigma_{\alpha_i}, \mu_{\beta_i}, \sigma_{\beta_i})\leftarrow(\mu_{\alpha_i}^0, \sigma_{\alpha_i}^0, \mu_{\beta_i}^0, \sigma_{\beta_i}^0), i =1,2,\cdots, p.$$ 
       \vspace{-1.7em}
  \FOR {$t=1,..,N_{\text{iter}}$}
      \STATE Sample batch $B$ of size $M$:
    \vspace{-0.6em}
      $$\alpha_i^j\sim \mathcal N(\mu_{\alpha_i}, \sigma_{\alpha_i}), \beta_i^j\sim \mathcal N(\mu_{\beta_i}, \sigma_{\beta_i}), \; i =1,2,\cdots, p, \ j = 1, 2, \cdots, M.$$
     \vspace{-1.5em}
      \STATE Compute the instantaneous fidelity and the averaged fidelity
        \vspace{-1.5em}
      $$F_j = \left\vert \mel{\psi_\ast}{U(\{\alpha_i^j, \beta_i^j\}_{i=1}^p)}{\psi_i} \right\vert^2, \quad \bar F = \frac{1}{M}\sum_{j=1}^M F_j.$$
        \vspace{-1.5em}
      \STATE Compute the policy gradient
      \[ \nabla_{\params}J(\params)= \frac{1}{M}\sum_{\{\alpha_i^j, \beta_i^j\}_{i=1}^p \in B } \nabla_\params 
      \log \pi_\params(\{\alpha_i^j, \beta_i^j\}_{i=1}^p)\cdot (F_j - \bar F). \] 
      \STATE Update weights $\params\leftarrow\params + \tau_t\nabla_{\params}J(\params)$.
  \ENDFOR
  \end{algorithmic}
  \vspace{0.5em}
\end{algorithm}

\agentname{} can be naturally extended to the setting of robust optimization for the Hamiltonian noise.  For the  max-min problem, the policy gradient in \autoref{eqn:pg} becomes

\begin{equation}
\nabla_{\params}J(\params) = 
\mathop{ \mathbb E}_{\substack{\alpha_i\sim \mathcal N(\mu_{\alpha_i}, \sigma_{\alpha_i}) \\ \beta_i\sim \mathcal N(\mu_{\beta_i}, \sigma_{\beta_i}) }}
\left[\nabla_{\params}\log \pi_{\params}(\{\alpha_i, \beta_i\}_{i=1}^p)
\cdot
 \min\limits_\delta  F(\{\alpha_i, \beta_i\}_{i=1}^p, \delta)
\right].
\label{eqn:pg-robust1}
\end{equation}

In practice, we sample independent random realizations $\delta_j$ from the noise region $\Delta$ at uniform, and use $\min\limits_j  F(\{\alpha_i, \beta_i\}_{i=1}^p, \delta_j)$ as an approximation in \autoref{eqn:pg-robust1}.
 When the fidelity itself is noisy such as the case of the Gaussian noise and the quantum noise, we simply use the measured fidelity in \autoref{eqn:gsfid} and \autoref{eqn:qfid} in the policy gradient step of \autoref{eqn:pg-robust1}.

\section{Quantum Qubit Models}
\label{sect:models}

We investigate the performance of \agentname{} for a single-qubit system, and two different multi-qubit systems, defined as follows:

\subsection{Single qubit model}
\label{set:sq_system}

Consider a single-qubit system, whose QAOA dynamics is generated by the Hamiltonians
\begin{equation}
  H_0= -\frac{1}{2}\sigma^z +2 \sigma^x,\qquad H_1= -\frac{1}{2}\sigma^z -2 \sigma^x,
\end{equation}
with $\sigma^\alpha$ the Pauli matrices.
The initial $\ket{\psi_i}$ and target $\ket{\psi_\ast}$ states are chosen to be the ground states of $H_i =  -\frac{1}{2}\sigma^z + \sigma^x$ and $H_\ast= -\frac{1}{2}\sigma^z - \sigma^x$, respectively. This control problem was introduced and analyzed in the context of reinforcement learning in Ref.~\citep{bukov2018reinforcement}: below the quantum speed limit (QSL), i.e. for total duration $T\leq T_\mathrm{QSL}\approx 2.41$, it is not possible to prepare the target state with unit fidelity; yet, in this regime there is a unique optimal solution which maximizes the fidelity of being in the target state, and its fidelity is less than $1$. Above the QSL, $T>T_\mathrm{QSL}$, there exist multiple unit-fidelity solutions to this constrained optimization problem.

\subsection{Multi-qubit Models}
\label{set:multi-qubits}

To compare the performance of \agentname{} against alternative algorithms, we use multi-qubit systems. For the purpose of a more comprehensive analysis, we use two different models, discussed in \citep{bukov2018reinforcement,niu2019optimizing}.

\subsubsection{Multi-qubit system I}
\label{set:mq1}

Consider first the transverse-field Ising model, described by the Hamiltonian \citep{bukov2018reinforcement}: 
\begin{equation}
    H[h]= -\sum_{j=1}^{N-1}\sigma_{j+1}^{z} \sigma_{j}^{z} -\sum_{j=1}^{N}(\sigma_{j}^{z} +  h \sigma_{j}^{x}).
    \label{eqn:mq1}
\end{equation}
Here $N$ is the total number of qubits.
The global control field $h\in\{\pm 4\}$ can take two discrete values, corresponding to the two alternating QAOA generators $H_0=H[-4]$ and $H_1=H[+4]$, cf.~\autoref{eqn:mq1}. The initial state $\ket{\psi_i}$ is the ground state of $H[-2]$, and the target state $\ket{\psi_\ast}$ is chosen to be the ground state of $H[+2]$, so the adiabatic regime is not immediately obvious; both states exhibit paramagnetic correlations and area-law bipartite entanglement. The overlap between the initial and target states goes down exponentially with increasing the number of qubits $N$ (with all other parameters kept fixed). This state preparation problem is motivated by applications in condensed matter theory.
For $N>2$, this qubit control problem was recently shown to exhibit similarities with optimization in glassy landscapes~\citep{day2019glassy}; for $N=2$ there exist durations $T$ for which the optimal solution is doubly-degenerate and the optimization landscape features symmetry-breaking~\citep{bukov2018broken}.

Additionally, we can also turn on small random Hamiltonian  noise to the interaction terms on the first two bonds of the spin system, denoted by $\omega_{1,2}$, which would mimic gate imperfections in the context of quantum computing:
\begin{equation}
     H[h;\omega_{1}, \omega_{2}]=-\left(1+\omega_{1}\right) \sigma_{1}^{z} \sigma_{2}^{z}-\left(1+\omega_{2}\right) \sigma_{2}^{z} \sigma_{3}^{z} -\sum_{j=3}^{N-1} \sigma_{j}^{z} \sigma_{j+1}^{z}-\sum_{j=1}^{N}\left(\sigma_{j}^{z}+h \sigma_{j}^{x}\right)
     \label{eqn:ns-mq1}
\end{equation}
The choice of noisy bonds is arbitrary. To keep the notation compact, we define the noise tuple $\delta=(\omega_1, \omega_2)$. Each $\omega_i\sim \mathrm{uniform}(\Delta)$ with $\Delta$ the support of the uniform distribution.

\subsubsection{Multi-qubit system II}
\label{set:mq2}
Consider another benchmark example \citep{niu2019optimizing}. Here, we choose the two alternating Hamiltonians from QAOA as 
\begin{equation}
    H_0 = \frac{1}{2}\left(\sigma_{N}^{z}+I_{N}\right), \quad H_1 = \sum_{i=1}^{N-1}\left(\sigma_{i}^{x} \sigma_{i+1}^{x}+\sigma_{i}^{y} \sigma_{i+1}^{y}\right),
    \label{eqn:mq2}
\end{equation}
where $I_N$ is the identity operator. 
The initial state is the product state $\ket{\psi_i}=|\overline{1}\rangle=|1\rangle_{1}|0\rangle_{2} \cdots|0\rangle_{N}$, and the target state is the product state $\ket{\psi_\ast}=|\overline{N}\rangle=|0\rangle_{1}|0\rangle_{2} \cdots|1\rangle_{N}$. This population transfer problem amounts to a qubit transfer.

The noisy multi-qubit system II uses the gate Hamiltonians: 
\begin{equation}
    H_0 = \frac{1}{2}\left(\sigma_{N}^{z}+I_{N}\right), \quad H_1(\delta) = \sum_{i=1}^{N-1}\left(\sigma_{i}^{x} \sigma_{i+1}^{x}+\sigma_{i}^{y} \sigma_{i+1}^{y}\right)  +\delta \sigma_{\left[\frac{N}{2}\right]-1}^{z} \sigma_{\left[\frac{N}{2}\right]}^{x} \sigma_{\left\lfloor\frac{N}{2}\right\rfloor+ 1}^{z},
    \label{eqn:ns-mq2}
\end{equation}
with $\delta \sim\text{uniform}(\Delta)$.
Here, the three-body noise term breaks the particle number (a.k.a.~magnetization) symmetry of the original noise-free system. 

\section{Numerical Experiments and Results}
\label{sect:errmitigate}

The models we introduced in \autoref{sect:models} were also studied in \citep{dong2019robust} using SCP to mitigate the error due to Hamiltonian noise. First, we benchmark our results against the derivative-based algorithms SCP and b-GRAPE \citep{wu2019learning} in the context of the Hamiltonian noise. We also present results for \agentname{} in the context of the Gaussian noise and quantum measurement noise. Then we compare our results to other derivative-free optimization methods, including Nelder-Mead \citep{gao2012implementing}, Powell \citep{powell1964efficient}, covariance matrix adaptation (CMA) \citep{hansen2001completely}, and particle swarm optimization (PSO) \citep{shi2001particle}.

All numerical experiments are performed on the Savio2 cluster at Berkeley Research Computing (BRC). Each node is equipped with Intel Xeon E5-2680 v4 CPUs with 28 cores. The \agentname{} is implemented in the TensorFlow 1.14 \citep{abadi2015tensorflow} along with TensorlFlow Probability 0.7.0 \citep{dillon2017tensorflow}. The quantum Hamiltonian environment is implemented using QuSpin \citep{weinberg2017quspin, weinberg2019quspin} and QuTIP \citep{johansson2012qutip, johansson2013qutip}. The two blackbox optimization methods CMA and PSO are implemented with Nevergrad \citep{nevergrad}. 

Throughout, we used the Adam optimizer \citep{kingma2014adam} to train \agentname{} with learning rate $10^{-2}$, and learning rate decay of $0.96$ applied every $50$ iteration steps. The training batch size $M$ is chosen either $128$ or $2048$ (see figure captions). The initial values for the standard deviation parameters of the policy, $\sigma_{\alpha_i}^{(0)}, \sigma_{\beta_i}^{(0)}, i =1,2,\cdots, p$, are either set to $0.0024$ or sampled from truncation log normal distribution with mean ${-3.0}$ and standard deviation $0.1$. In the Single-qubit testcase (cf.~\autoref{set:sq}) and the Multi-qubit I testcase (cf.~\autoref{set:mq1}), the initial values for the mean parameters of the policy, $\mu_{\alpha_i}^{(0)}, \mu_{\beta_i}^{(0)}, i =1,2,\cdots, p$, are randomly sampled from a truncated normal distribution with mean $0.5$ and standard deviation $0.1$. In the Multi-qubit II testcase (cf.~\autoref{set:mq2}) , the initial values for the means are sampled from a truncated normal distribution with mean $3.0$ and standard deviation $0.1$. In practice, we noticed that the performance of \agentname{} is sensitive to the initialization of the means $\mu_{\alpha_i},\mu_{\beta_i}$. In some cases, the initialization was tuned to achieve better performance (c.f.~\autoref{fig:mq2}).  

In the numerical experiments, we do not enforce hard constraints on the positivity of $\alpha_i$ and $\beta_i$; yet, in practice we were still able to obtain protocols with positive $\alpha_i\geq 0$ and $\beta_i\geq 0$. This is mainly because the initialization of the mean parameters in the policy is positive and sufficiently far away from zero, and because there are already optimal protocol solutions (i.e.~local minima of the control landscape close to the initial values) with positive $\alpha_i$ and $\beta_i$.  
 
\subsection{Single qubit results}
\label{set:sq}

\begin{figure}[htbp]
    \centerline{\includegraphics[scale=0.38]{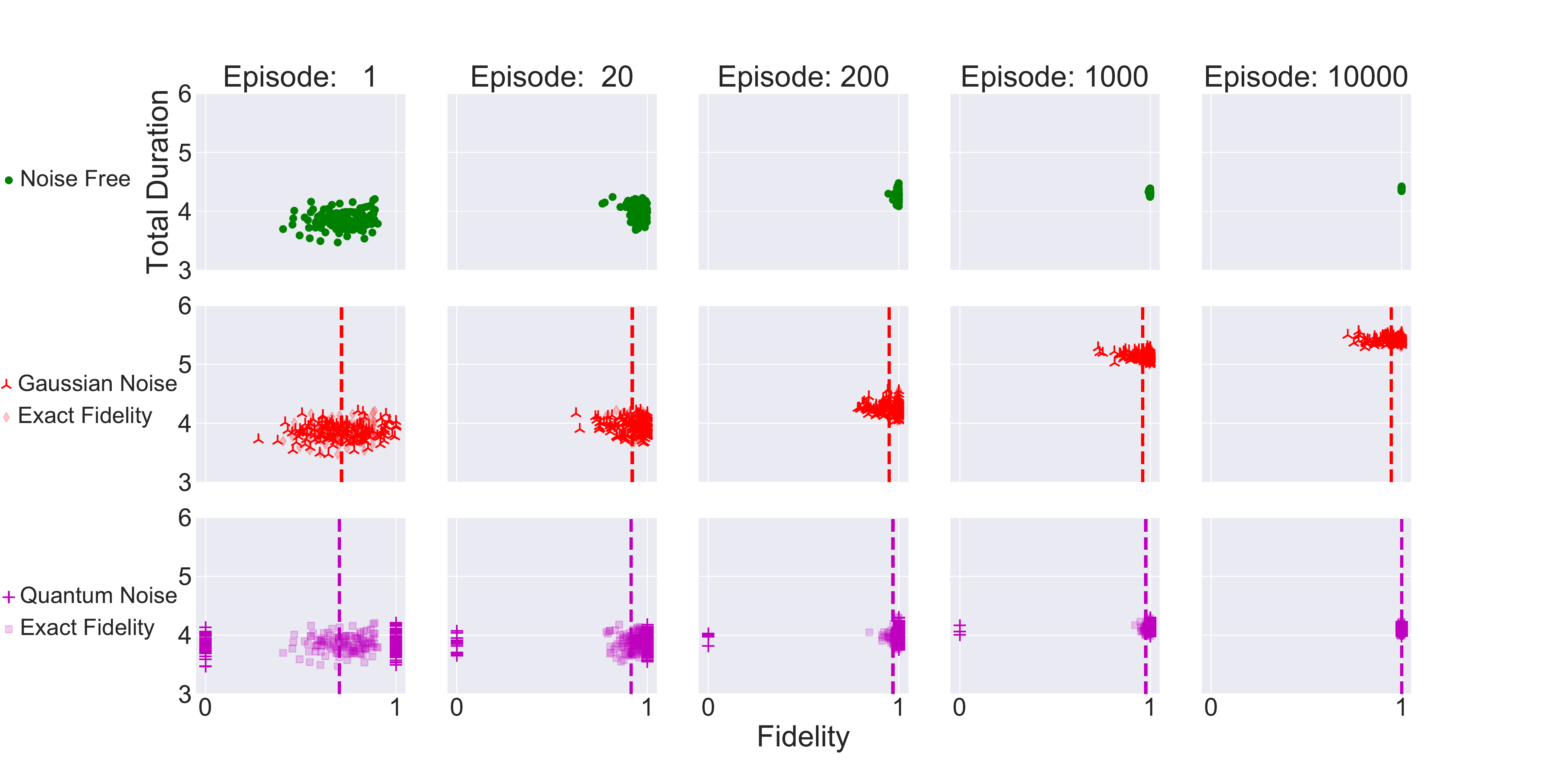}}
    \caption{\small The distribution in the learning process for the single-qubit testcase. From left to right, snapshots of the training batch distribution in the (protocol duration, fidelity) space at different training episodes for \agentname{}. Top row: noise-free fidelity problem (green circles).
    Middle row: Gaussian fidelity noise problem (red tri-ups), the corresponding exact fidelity values for comparison only (red diamonds, not used in training), and the mean mini-batch fidelity (dashed vertical line).
    Bottom row: quantum measurement noise problem (magenta crosses) with binary values $\{0,1\}$, the corresponding exact fidelity values for comparison only (magenta squares, not used for training), and the mean mini-batch fidelity (dashed vertical line). 
    The final learned distributions represent a set of solutions with different total protocol durations but still sharing the same optimal fidelity, demonstrating the machine learning aspect of the algorithm (see text). 
     The standard deviation of the Gaussian noise is $0.1$. The QAOA depth is $p=4$. The \agentname{} algorithm is trained with a single minibatch of size $M=128$ for a total iteration number $N_{\text{iter}}=10^4$. The initial mean values  $\mu_{\alpha_i}^{(0)}, \mu_{\beta_i}^{(0)}$ are randomly sampled from a truncated normal distribution with mean $0.5$ and standard deviation $0.1$ (i.e. $\mathcal N(0.5, 0.1^2)$) and the initial standard deviation values $\sigma_{\alpha_i}^{(0)}, \sigma_{\beta_i}^{(0)}$ are sampled from a truncated log normal distribution of mean $-3.0$ and standard deviation $0.1$ (i.e. $\operatorname{Lognormal}(-3.0, 0.1^2)$).   } 
    \label{fig:nonoise-sq}
\end{figure}

Figure~\ref{fig:nonoise-sq} (topmost row) shows snapshots of the policy during training for \agentname{} in the noise-free case. We sample a batch of protocols from the policy learned in the middle of training, and show its distribution in (protocol duration, fidelity)-space. Due to the random initialization of the policy parameters $\boldsymbol{\theta}$, the algorithm starts from a broad distribution. After the number of training episodes (a.k.a.~optimization iterations) increases, the mean of the training batch distribution shifts ever closer to the unit-fidelity region, as expected. At the same time, the distribution also shrinks at later training episodes, and becomes approximately a delta-function in fidelity space in the infinite-training-episode limit since the environment for the noise-free problem is deterministic (though the distribution may still have exhibit finite width due to the decay of the learning rate in the optimization procedure). 

Figure~\ref{fig:nonoise-sq} (middle and bottom rows) shows the effect of the two types  of noise on the performance of \agentname{}. We test both the \emph{Gaussian noise}, which takes into account various classical potential measurement uncertainty sources in the lab, as well as the intrinsic \emph{quantum measurement noise}  induced by collapsing the wavefunction during measurements. In the case of quantum measurement noise (magenta), we use only binary fidelity values for the reward \agentname{}, cf.~Eq.~(\ref{eqn:qfid}); the exact fidelity values for the batch (which are not binary) are shown for comparison purposes only. We emphasize that we do not repeat the quantum measurement on the \emph{same} protocol several times, but only take a single quantum measurement for each protocol from the sampled batch in every iteration. The mean batch fidelity is shown as a vertical dashed line. In the case of Gaussian noise (red), the noisy fidelity values used for training are  not binary; \agentname{} is thus well-suited to handle both classical and quantum noise effects. Because we clip the Gaussian-noisy fidelities to fit in the interval $[0,1]$, the mean fidelity of the policy (vertical dashed red line) remains slightly away from unity even after a large number of training episodes, introducing a small gap, also visible in the training curves for the multi-qubit examples (\autoref{fig:ns_q_fid_curve}, left).

Note that the policy optimized using \agentname{} converges at later training episodes for both noisy settings (measurement and Hamiltonian noise). An interesting  feature is the remaining finite width along the protocol duration axis: these unit-fidelity protocols are indistinguishable from the point of view of the objective function and are thus equally optimal. Hence, above the QSL, \agentname{} is capable of learning multiple solutions simultaneously, unlike conventional optimal control algorithms, showcasing one of the advantages of using reinforcement learning. We can indeed verify that these distribution points correspond to distinct protocols, by visualizing the batch trajectories on the Bloch sphere (the projective space of the single-qubit Hilbert space), cf.~\autoref{fig:trace-sq}. We mention in passing that, depending on the initialization of the policy parameters, \agentname{} finds a different (but equivalent w.r.t.~the reward) local basin of attraction in the control landscape, as can be seen from the difference in the mean total protocol duration at later training episodes for the noise-free and the two noisy cases.

\subsection{Multi-qubit results}

\begin{figure}[htbp]
    \centerline{
        \subfigure[Multi-qubit I]{
      \includegraphics[scale=0.26]{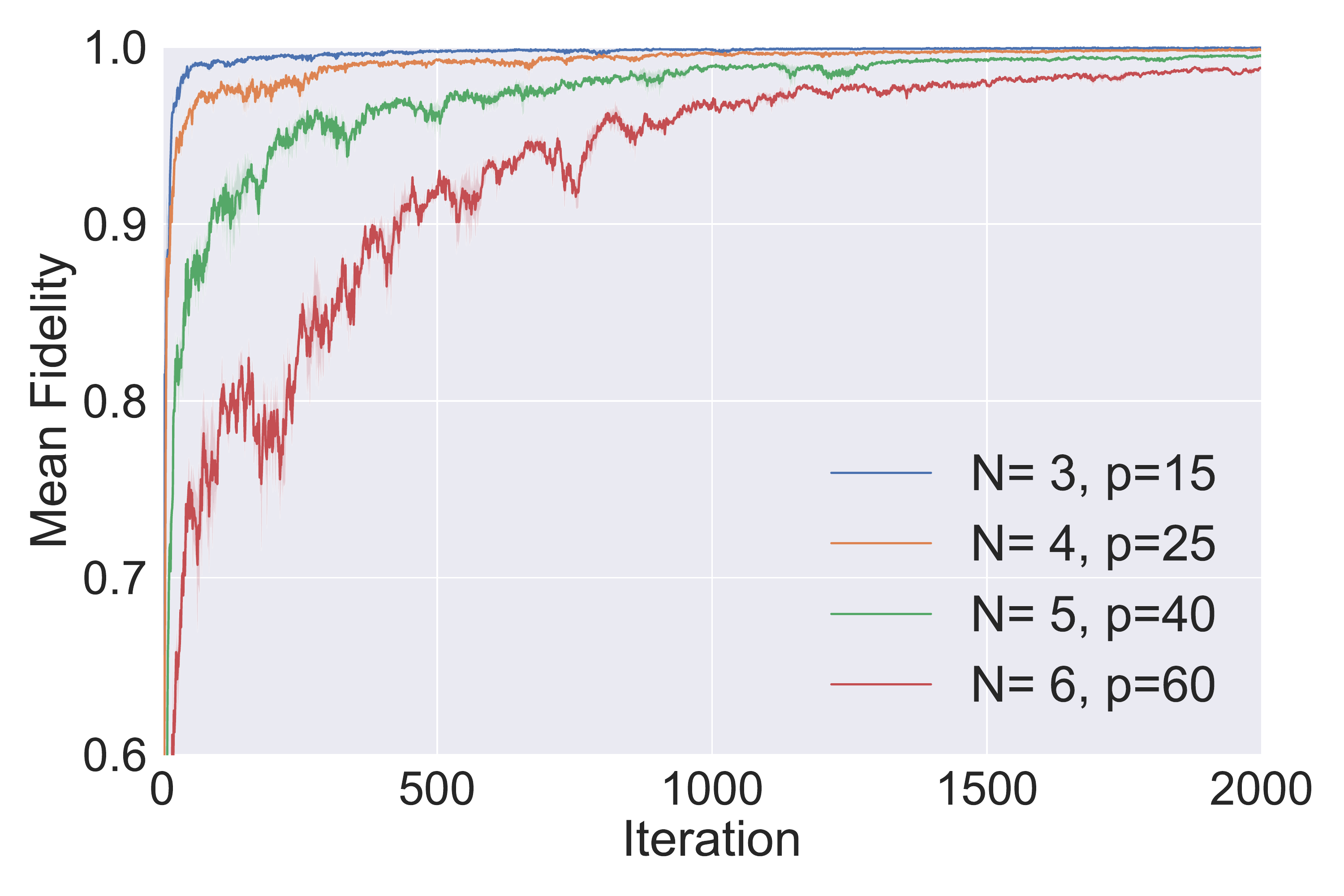}
            }
           \subfigure[Multi-qubit II]{
        \includegraphics[scale=0.26]{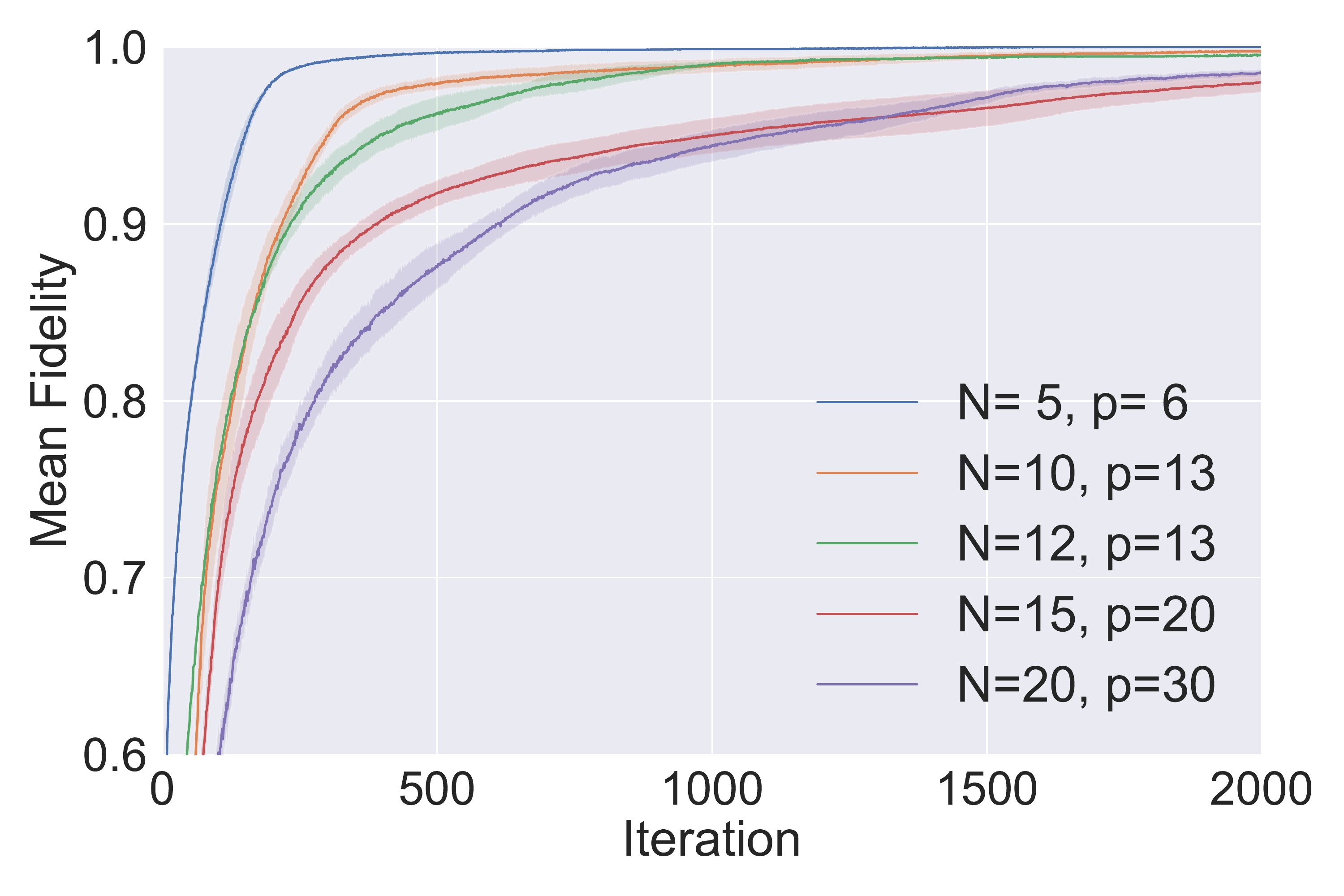}
        }}
    \caption{\small Multi-qubit systems, noise-free case. Learning curves (reward vs.~episode number) for the Multi-qubit I testcase (a) and the Multi-qubit II testcase (b), for a different number of qubits $N$ and QAOA depth $p$ for three different random seeds. The \agentname{} algorithm is trained with batch size $M$ of 128 for 2000 iterations. The means initialization is sampled from truncated $\mathcal N(0.5, 0.1^2)$ [left] and  $\mathcal N(1.5, 0.1^2)$ [right]. The stds initialization is from truncated $\operatorname{Lognormal}(-3.0, 0.1^2)$.  
    }
        \label{fig:nonoise-mq}
        \vspace{-4mm}
\end{figure}

Figure~\ref{fig:nonoise-mq} shows the training curves of \agentname{} for an increasing number of qubits $N$ and QAOA depths $p$. In accord with the fact that the multi-spin fidelity decreases exponentially with increasing $N$, the \agentname{} algorithm takes longer to converge. 

\begin{figure}[htbp]
    \centerline{
        \includegraphics[scale=0.26]{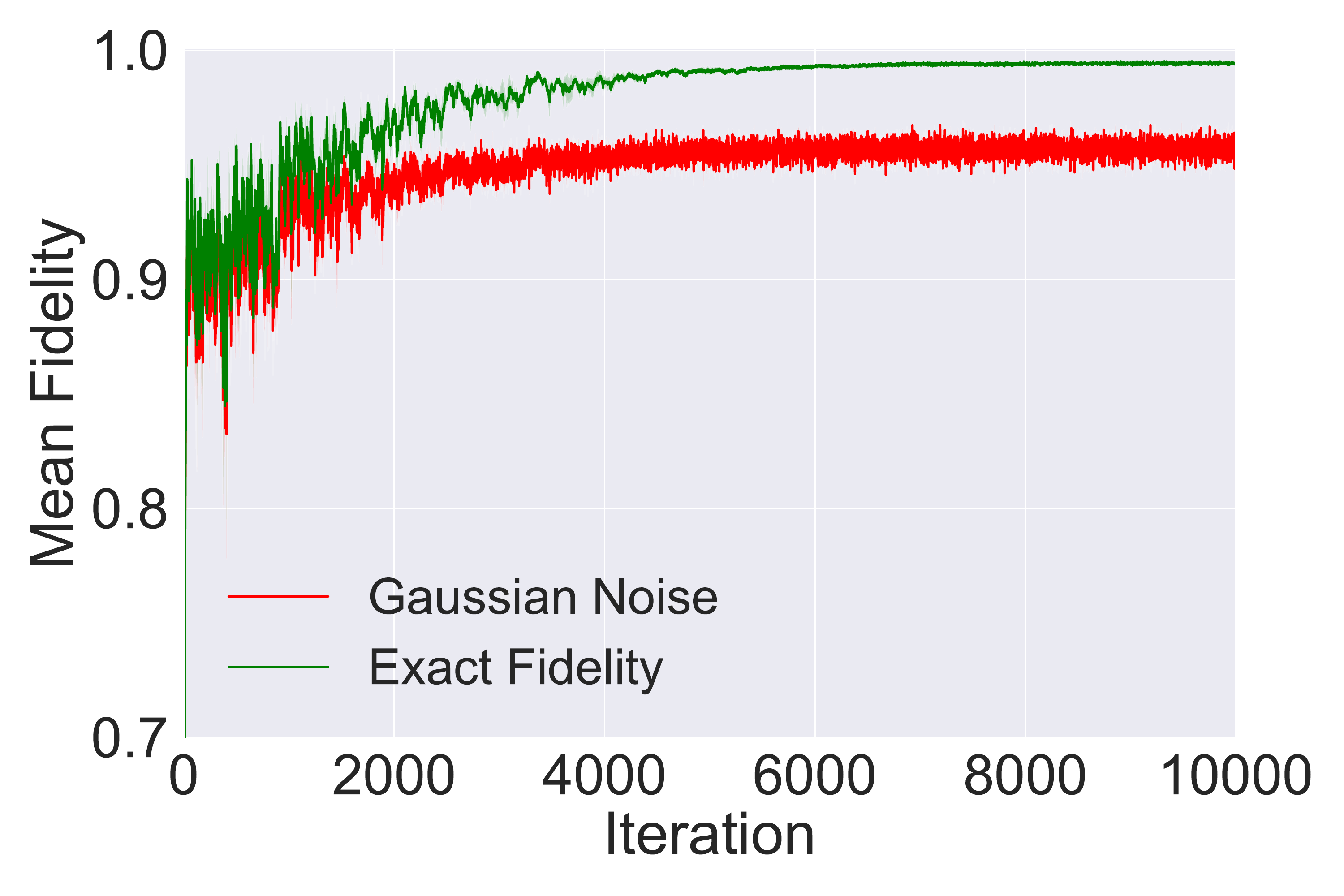}
        \includegraphics[scale=0.26]{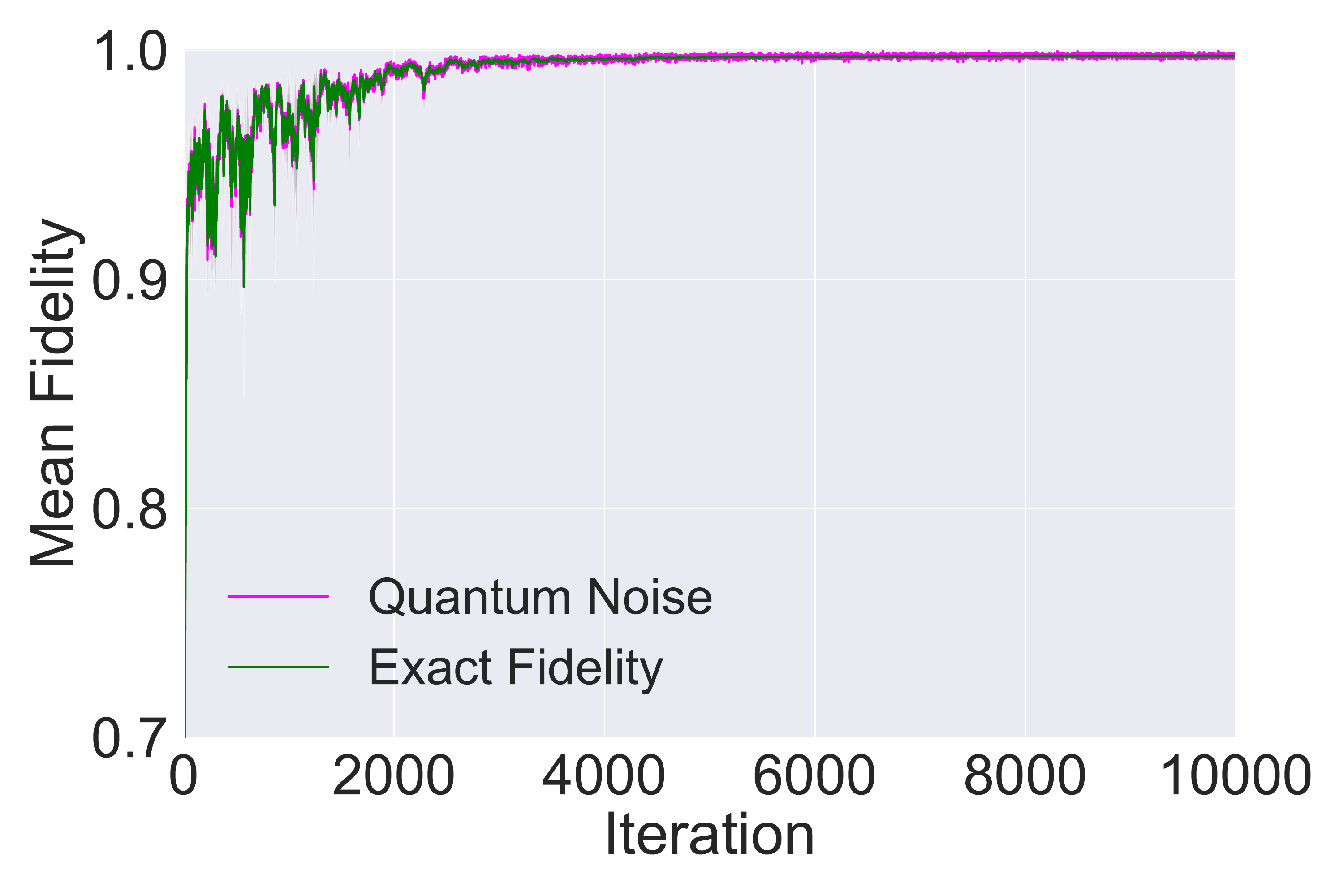}}
        \caption{\small Multi-qubit testcase I, training curves: the reward (mean batch fidelity, red for Gaussian noise and magenta for quantum measure noise) used in \agentname{} against the number of training episodes (i.e.~iterations). For comparison purposes only, we also show the exact noise-free mean fidelity (green). Left: Gaussian noise. Right: quantum measurement noise. 
        The standard deviation of the Gaussian noise is $0.1$. The number of qubits is $N=3$. The batch sizes for Gaussian noise and quantum measurement noise are $128$ and $2048$, respectively.  The initial mean values are sampled from truncated $\mathcal N(0.5, 0.1^2)$ and the initial standard deviation values -- from truncated $\operatorname{Lognormal}(-3.0, 0.1^2)$ for both noisy cases.  
        }
        \label{fig:ns_q_fid_curve}
        \vspace{-0.5cm}
 \end{figure}

Adding Gaussian and quantum measurement noise, in Fig.~\ref{fig:ns_q_fid_curve} we show the training curves for \agentname{} for $N\!=\!3$ qubits. For each noisy case, we present the actual mean fidelities (red for the Gaussian noise and magenta for the quantum measurement noise); the exact fidelities (green) are shown only for comparison and are not used in training. Note that learning from quantum measurements is more prone to noise in the initial stage of the optimization, yet the algorithm converges within a smaller number of episodes compared to the case of the Gaussian noise. For Gaussian noise, similar to the single-qubit case, we observe a small gap between the exact fidelity and the noisy fidelity due to clipping the noisy fidelities to fit within the interval $[0,1]$. Empirically, we observe the gap size to be almost always about half the Gaussian noise level. This indicates that the probability distribution is moving towards the correct direction (with fidelity close to unity) even though the observed fidelity is away from it due to the noise. More results are presented for the Gaussian noise and for the quantum measurement noise in \autoref{sec:app-gsfid} and \autoref{sec:app-qfid} in the Appendix, respectively. In \autoref{fig:gsfid-plus}, the optimization becomes more difficult with increasing qubit number $N$ and the gap is proportionally enlarged according to the Gaussian noise level. In \autoref{fig:qfid-plus}, we show that the variance of the mean fidelities is reduced at larger batch sizes for the quantum measurement noise, and similar results can be observed for the Gaussian noise as well.

We now benchmark \agentname{} against a number of different optimal control algorithms.  
In order to compare \agentname{} with state-of-the-art optimization methods using gradient and Hessian information such as b-GRAPE and SCP, we evaluate their performance using both the batch average and the worst-case fidelity as reference. For protocol durations  $\{\alpha_i, \beta_i \}_{i=1}^p$, the average and worst-case fidelity within a given support for the uniform distribution $\Delta$, are defined as 

\begin{eqnarray}
    F_{\text{avg}} (\{\alpha_i, \beta_i\}_{i=1}^p) &=&\frac{1}{|\Delta|}  \int_\Delta  {F}(\{\alpha_i, \beta_i\}_{i=1}^p, \delta) \ \mathrm d\delta 
    \label{eqn:avg-fid}\\
        F_{\text{worst-case}} (\{\alpha_i, \beta_i\}_{i=1}^p) &=&\min_{\delta\in \Delta}  {F}(\{\alpha_i, \beta_i\}_{i=1}^p, \delta). 
    \label{eqn:wc-fid}
\end{eqnarray}

\begin{figure}[htbp]
    \centerline{
        \includegraphics[scale=0.42]{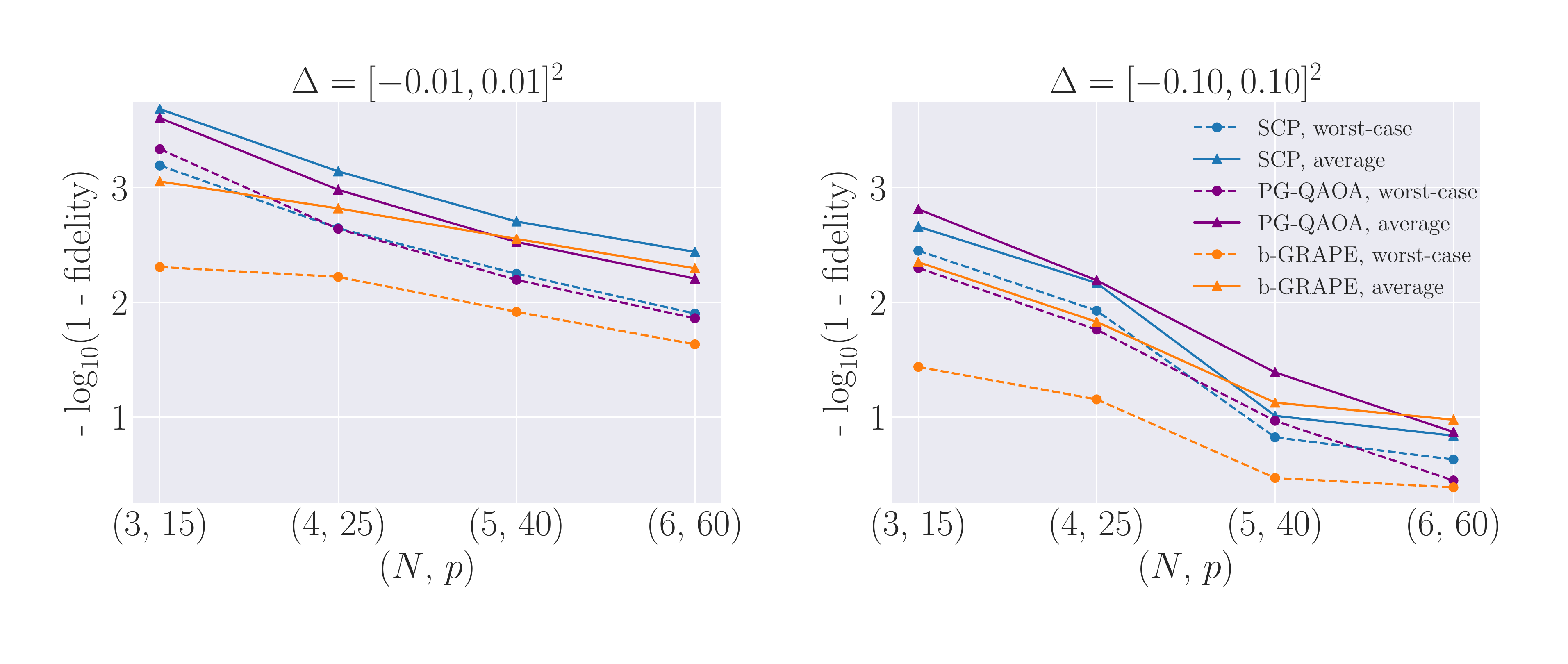}}
    \vspace{-1cm}
    \caption{\small Multi-qubit testcase I, algorithms comparison for Hamiltonian gate noise. Fidelity achieved by \agentname{} (purple), SCP (blue) and b-GRAPE (orange) for a few different numbers of qubits $N$ and total QAOA depth values $p$.
    The two panels correspond to different values of the support $\Delta$ of the uniform distribution used for the Hamiltonian gate noise.
    We show both the average fidelity (solid lines), and the worst protocol (dashed lines), cf.~\autoref{eqn:avg-fid} and \autoref{eqn:wc-fid}, respectively. The \agentname{} algorithm is trained with the mini-batch size $M=128$, except $N=6$, where $M=1024$. The initial values for the means are sampled from a truncated $\mathcal N(0.5, 0.1^2)$ and the initial values for the standard deviations were kept constant at $0.0024$. 
    }
        \label{fig:mq1}
        \vspace{-0.5cm}
 \end{figure}

\begin{figure}[H]
    \centering
        \includegraphics[width=0.6\textwidth]{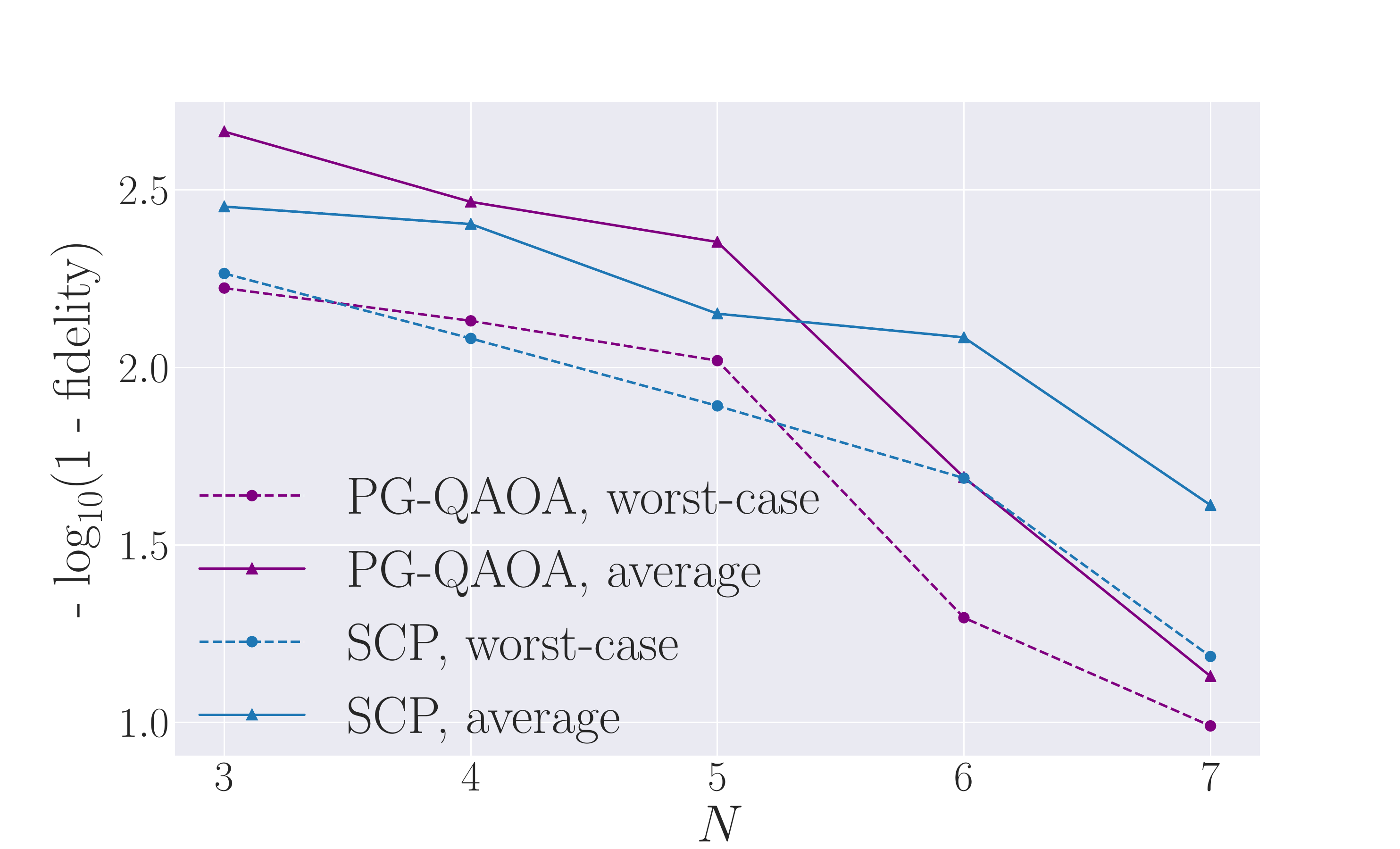}
    \caption{\small Multi-qubit testcase II, algorithms comparison for Hamiltonian gate noise. The comparison between \agentname{} (purple) and SCP (blue) in terms of robust QAOA for multi-qubit case II with different number of qubits $N$. The QAOA depth is $p\!=\!N+1$ and the support $\Delta$ of the uniform distribution used for the Hamiltonian gate noise is $[-0.15, 0.15]$. We show both the average fidelity (solid lines), and the worst protocol (dashed lines), cf.~\autoref{eqn:avg-fid} and \autoref{eqn:wc-fid}, respectively. The \agentname{} algorithm is trained with minibatch sizes $M=128$ for $10^4$ iterations. The initial values of the standard deviations are kept constant at $0.0024$; the initial values for the means were drawn from $\mathcal N(1.0, 0.2^2)$ for $N=3$, $\mathcal N(1.5, 0.2^2)$ for $N=4$, and $\mathcal N(3.0, 0.2^2)$ for $N>4$. }
        \label{fig:mq2}
        \vspace{-0.5cm}
 \end{figure}

A comparison for testcases multi-qubit I and II are shown in \autoref{fig:mq1} and \autoref{fig:mq2}, respectively. In terms of both the average and worst case, \agentname{} performs comparably to the SCP; although \agentname{} is derivative-free and uses a first-order derivative optimizer, it can occasionally even reach better solutions than SCP w.r.t.~the average fidelity. \agentname{} clearly outperforms b-GRAPE \citep{wu2019learning} in the numerical experiments involving a small number of qubits.
We also observe a performance drop for \agentname{} when the number of qubits is increased. Properly scaling up the performance of \agentname{} with increasing $N$ remains a topic of further investigation.

\begin{figure}[h]
    \centerline{
    \includegraphics[scale=0.41]{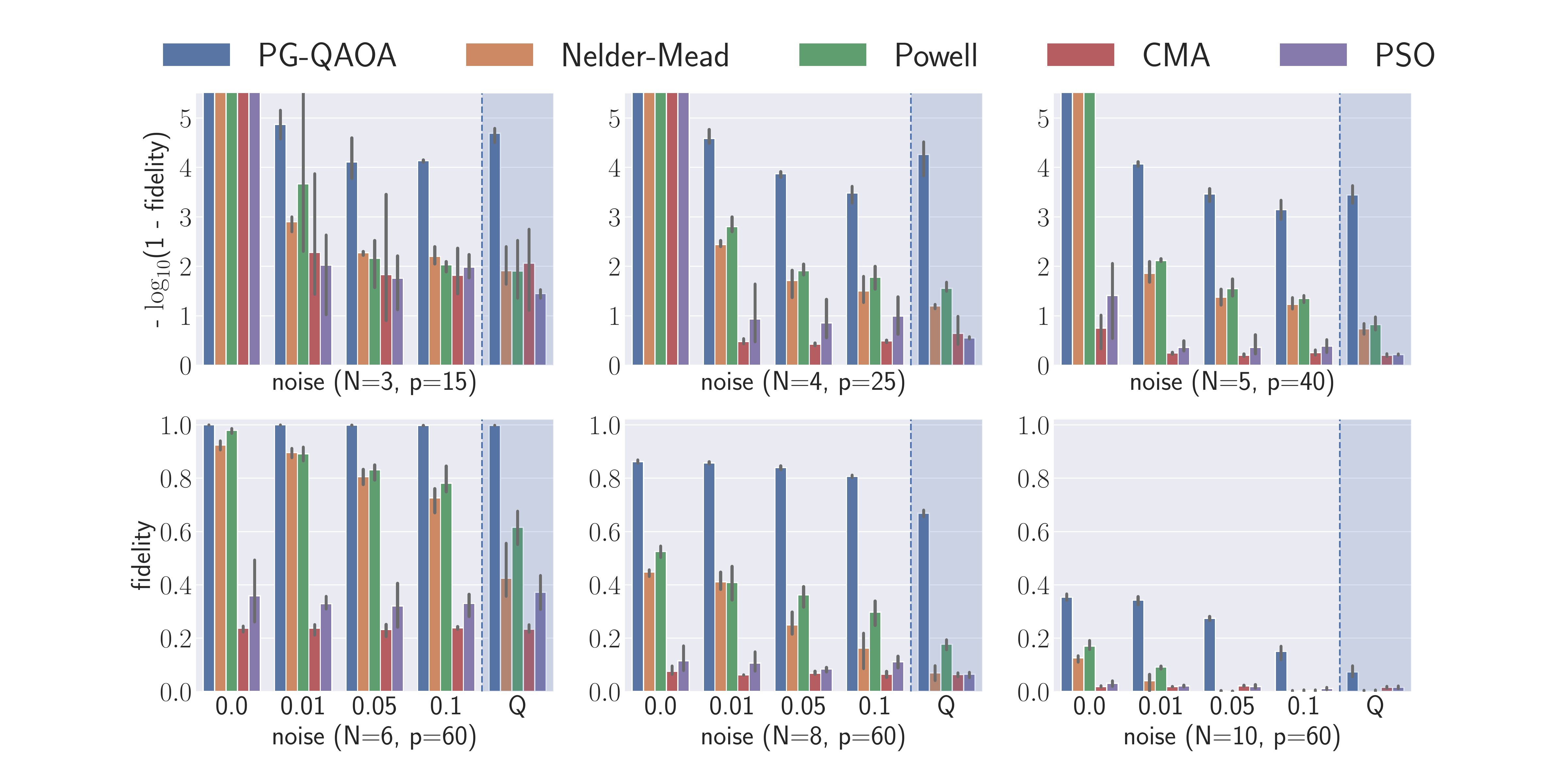}}  
    \caption{\small Multi-qubit testcase I. Comparison between different optimization algorithms for $N=3,4,5$ qubits (the first row) and $N=6,8,10$ qubits (the second row), and different fidelity noise level (cf.~$x$-axis for the standard deviation of the Gaussian noise; the label "Q" (shaded area) stands for quantum measurement noise): \agentname{} (blue), Nelder-Mead (orange), Powell (green), CMA (red), and PSO (purple). The comparison is in log-scale (upper row), and the normal scale (lower row). \agentname{} outperforms the rest in the presence of noise. The batch sizes are $M=2048$ for all the methods, except for $N=10$, where $M=256$, and the total number of iterations is $10^4$.  For all \agentname{} experiments, the initial values for the means are sampled from a truncated $\mathcal N(0.5, 0.1^2)$ and the standard deviations initialization -- from truncated $\operatorname{Lognormal}(-3.0, 0.1^2)$.
    }
        \label{fig:bo_comp}
        \vspace{-0.5cm}
 \end{figure}

Last, in \autoref{fig:bo_comp} we show the comparison among other widely used blackbox optimization methods, such as Nelder-Mead, Powell, covariance matrix adaptation (CMA) and particle swarm optimization (PSO). In contrast to \agentname{} which learns in distribution (i.e.~in practice using MC-sampled batches), the other algorithms accept a single scalar cost function value to optimize. Therefore, we use the mean fidelity over a (potentially noisy) training batch; this constitutes a fair comparison, since the mean batch fidelity is precisely the definition of the reward in policy gradient. The different algorithms have a comparable performance in the noise-free case (\autoref{fig:bo_comp}, leftmost column). In the presence of measurement noise in the reward function, we observe a decrease in performance in all algorithms. At the same time, \agentname{} still outperforms other algorithms, which is clearly visible when the number of qubits $N$ increases~\footnote{Note that, for $N\!=\!6,8,10$, we keep $p\!=\!60$ fixed, so the maximum obtainable fidelity is expected to decrease.}. \agentname{} appears less sensitive to the size of the Gaussian noise; moreover, \agentname{} appears particularly suitable for handling the quantum measurement noise.

\section{Conclusion and Outlook}
\label{sect:sum}
Due to intrinsic limitation of near term quantum devices, error mitigation techniques can be essential for the performance of quantum variational algorithms such as QAOA. Many classical optimization algorithms (derivative-free or those requiring derivative information) may not perform well in the presence of noise. We demonstrate that probability-based optimization methods from reinforcement learning can be well suited for such tasks. This work considers the simplest setup, where we parameterize each optimization variable using only two variables describing an i.i.d.~Gaussian distribution. The probability distribution is then optimized using the policy gradient method, which allows to handle continuous control problems. We demonstrate that PG-QAOA does not require derivatives to be computed explicitly, and can perform well even if the objective function is not smooth with respect to the error. The performance of PG-QAOA may even be sometimes comparable to that of much more sophisticated algorithms, such as  sequential convex programming (SCP), which require information of first and second order derivatives of the objective function. PG-QAOA also compares favorably to a number of commonly used blackbox optimization methods, particularly in experiments with noise and other sources of uncertainty. 

Viewed from the perspective of reinforcement learning, the Gaussian probability distribution used in this work is one of the simplest possible choices. More involved distributions, such as multi-modal Gaussian distributions, normalizing flow-based models~\citep{kingma2016improved, dinh2016density}, autoregressive models~\citep{germain2015made},and long short-term memory (LSTM) models may be considered. Based on our preliminary results, these methods can introduce a significantly larger number of parameters, but the benefit is not yet obvious. We can also employ more advanced RL algorithms, such as the natural policy gradient method (NPG)~\citep{kakade2002natural}, the trust region policy optimization (TRPO)~\citep{schulman2015trust} and the proximal policy optimization method (PPO)~\citep{schulman2017proximal}. Finally, this work only considers implementations on a classical computer. Implementing and testing PG-QAOA on near term quantum computing devices such as those provided by IBM Q will be our future work.

\acks{This work was partially supported by a Google Quantum Research Award (L.L., J.Y.) and by the Department of Energy under Grant No. DE-AC02-05CH11231 and No. DE-SC0017867 (L.L.). M.B.~was supported by the Emergent Phenomena in Quantum Systems initiative of the Gordon and Betty Moore Foundation, and the U.S. Department of Energy, Office of Science, Office of Advanced Scientific Computing Research, Quantum Algorithm Teams Program. We thank Yulong Dong for helpful discussions, and Berkeley Research Computing (BRC) for providing computational resources.}

\clearpage
\bibliography{msml2020}

\begin{thebibliography}{62}
\providecommand{\natexlab}[1]{#1}
\providecommand{\url}[1]{\texttt{#1}}
\expandafter\ifx\csname urlstyle\endcsname\relax
  \providecommand{\doi}[1]{doi: #1}\else
  \providecommand{\doi}{doi: \begingroup \urlstyle{rm}\Url}\fi

\bibitem[Abadi et~al.(2015)Abadi, Agarwal, Barham, Brevdo, Chen, Citro,
  Corrado, Davis, Dean, Devin, et~al.]{abadi2015tensorflow}
Mart{\i}n Abadi, Ashish Agarwal, Paul Barham, Eugene Brevdo, Zhifeng Chen,
  Craig Citro, Greg~S Corrado, Andy Davis, Jeffrey Dean, Matthieu Devin, et~al.
\newblock Tensorflow: Large-scale machine learning on heterogeneous systems,
  2015.
\newblock \emph{Software available from tensorflow. org}, 1\penalty0 (2), 2015.

\bibitem[An and Zhou(2019)]{an2019deep}
Zheng An and DL~Zhou.
\newblock Deep reinforcement learning for quantum gate control.
\newblock \emph{arXiv preprint arXiv:1902.08418}, 2019.

\bibitem[August and Hern{\'a}ndez-Lobato(2018)]{august2018taking}
Moritz August and Jos{\'e}~Miguel Hern{\'a}ndez-Lobato.
\newblock Taking gradients through experiments: Lstms and memory proximal
  policy optimization for black-box quantum control.
\newblock In \emph{International Conference on High Performance Computing},
  pages 591--613. Springer, 2018.

\bibitem[Brookes et~al.(2019)Brookes, Busia, Fannjiang, Murphy, and
  Listgarten]{brookes2019view}
David~H Brookes, Akosua Busia, Clara Fannjiang, Kevin Murphy, and Jennifer
  Listgarten.
\newblock A view of estimation of distribution algorithms through the lens of
  expectation-maximization.
\newblock \emph{arXiv preprint arXiv:1905.10474}, 2019.

\bibitem[Bukov(2018)]{bukov2018quantum}
Marin Bukov.
\newblock Reinforcement learning for autonomous preparation of
  floquet-engineered states: Inverting the quantum kapitza oscillator.
\newblock \emph{Physical Review B}, 98\penalty0 (22):\penalty0 224305, 2018.

\bibitem[Bukov et~al.(2018{\natexlab{a}})Bukov, Day, Sels, Weinberg,
  Polkovnikov, and Mehta]{bukov2018reinforcement}
Marin Bukov, Alexandre~GR Day, Dries Sels, Phillip Weinberg, Anatoli
  Polkovnikov, and Pankaj Mehta.
\newblock Reinforcement learning in different phases of quantum control.
\newblock \emph{Physical Review X}, 8\penalty0 (3):\penalty0 031086,
  2018{\natexlab{a}}.

\bibitem[Bukov et~al.(2018{\natexlab{b}})Bukov, Day, Weinberg, Polkovnikov,
  Mehta, and Sels]{bukov2018broken}
Marin Bukov, Alexandre~GR Day, Phillip Weinberg, Anatoli Polkovnikov, Pankaj
  Mehta, and Dries Sels.
\newblock Broken symmetry in a correlated quantum control landscape.
\newblock \emph{Physical Review A}, 2018{\natexlab{b}}.

\bibitem[Chen et~al.(2013)Chen, Dong, Li, Chu, and Tarn]{chen2013fidelity}
Chunlin Chen, Daoyi Dong, Han-Xiong Li, Jian Chu, and Tzyh-Jong Tarn.
\newblock Fidelity-based probabilistic q-learning for control of quantum
  systems.
\newblock \emph{IEEE transactions on neural networks and learning systems},
  25\penalty0 (5):\penalty0 920--933, 2013.

\bibitem[Chen and Xue(2019)]{chen2019manipulation}
Jun-Jie Chen and Ming Xue.
\newblock Manipulation of spin dynamics by deep reinforcement learning agent.
\newblock \emph{arXiv preprint arXiv:1901.08748}, 2019.

\bibitem[Dalgaard et~al.(2019)Dalgaard, Motzoi, Sorensen, and
  Sherson]{dalgaard2019global}
Mogens Dalgaard, Felix Motzoi, Jens~Jakob Sorensen, and Jacob Sherson.
\newblock Global optimization of quantum dynamics with alphazero deep
  exploration.
\newblock \emph{arXiv preprint arXiv:1907.05672}, 2019.

\bibitem[Day et~al.(2019)Day, Bukov, Weinberg, Mehta, and Sels]{day2019glassy}
Alexandre~GR Day, Marin Bukov, Phillip Weinberg, Pankaj Mehta, and Dries Sels.
\newblock Glassy phase of optimal quantum control.
\newblock \emph{Physical review letters}, 122\penalty0 (2):\penalty0 020601,
  2019.

\bibitem[Dillon et~al.(2017)Dillon, Langmore, Tran, Brevdo, Vasudevan, Moore,
  Patton, Alemi, Hoffman, and Saurous]{dillon2017tensorflow}
Joshua~V Dillon, Ian Langmore, Dustin Tran, Eugene Brevdo, Srinivas Vasudevan,
  Dave Moore, Brian Patton, Alex Alemi, Matt Hoffman, and Rif~A Saurous.
\newblock Tensorflow distributions.
\newblock \emph{arXiv preprint arXiv:1711.10604}, 2017.

\bibitem[Dinh et~al.(2016)Dinh, Sohl-Dickstein, and Bengio]{dinh2016density}
Laurent Dinh, Jascha Sohl-Dickstein, and Samy Bengio.
\newblock Density estimation using real nvp.
\newblock \emph{arXiv preprint arXiv:1605.08803}, 2016.

\bibitem[Dong et~al.(2019)Dong, Meng, Lin, Kosut, and Whaley]{dong2019robust}
Yulong Dong, Xiang Meng, Lin Lin, Robert Kosut, and K~Birgitta Whaley.
\newblock Robust control optimization for quantum approximate optimization
  algorithm.
\newblock \emph{arXiv preprint arXiv:1911.00789}, 2019.

\bibitem[Farhi et~al.(2014)Farhi, Goldstone, and Gutmann]{farhi2014quantum}
Edward Farhi, Jeffrey Goldstone, and Sam Gutmann.
\newblock A {Quantum Approximate Optimization Algorithm}.
\newblock \emph{arXiv preprint arXiv:1411.4028}, 2014.

\bibitem[F{\"o}sel et~al.(2018)F{\"o}sel, Tighineanu, Weiss, and
  Marquardt]{fosel2018reinforcement}
Thomas F{\"o}sel, Petru Tighineanu, Talitha Weiss, and Florian Marquardt.
\newblock Reinforcement learning with neural networks for quantum feedback.
\newblock \emph{Physical Review X}, 8\penalty0 (3):\penalty0 031084, 2018.

\bibitem[Gao and Han(2012)]{gao2012implementing}
Fuchang Gao and Lixing Han.
\newblock Implementing the nelder-mead simplex algorithm with adaptive
  parameters.
\newblock \emph{Computational Optimization and Applications}, 51\penalty0
  (1):\penalty0 259--277, 2012.

\bibitem[Germain et~al.(2015)Germain, Gregor, Murray, and
  Larochelle]{germain2015made}
Mathieu Germain, Karol Gregor, Iain Murray, and Hugo Larochelle.
\newblock Made: Masked autoencoder for distribution estimation.
\newblock In \emph{International Conference on Machine Learning}, pages
  881--889, 2015.

\bibitem[Greensmith et~al.(2004)Greensmith, Bartlett, and
  Baxter]{greensmith2004variance}
Evan Greensmith, Peter~L Bartlett, and Jonathan Baxter.
\newblock Variance reduction techniques for gradient estimates in reinforcement
  learning.
\newblock \emph{Journal of Machine Learning Research}, 5\penalty0
  (Nov):\penalty0 1471--1530, 2004.

\bibitem[Hadfield(2018)]{hadfield2018quantum}
Stuart Hadfield.
\newblock Quantum algorithms for scientific computing and approximate
  optimization.
\newblock \emph{arXiv preprint arXiv:1805.03265}, 2018.

\bibitem[Hadfield et~al.(2019)Hadfield, Wang, O’Gorman, Rieffel, Venturelli,
  and Biswas]{hadfield2019quantum}
Stuart Hadfield, Zhihui Wang, Bryan O’Gorman, Eleanor~G Rieffel, Davide
  Venturelli, and Rupak Biswas.
\newblock From the quantum approximate optimization algorithm to a quantum
  alternating operator ansatz.
\newblock \emph{Algorithms}, 12\penalty0 (2):\penalty0 34, 2019.

\bibitem[Hansen and Ostermeier(2001)]{hansen2001completely}
Nikolaus Hansen and Andreas Ostermeier.
\newblock Completely derandomized self-adaptation in evolution strategies.
\newblock \emph{Evolutionary computation}, 9\penalty0 (2):\penalty0 159--195,
  2001.

\bibitem[Ho and Hsieh(2019)]{ho2019efficient}
Wen~Wei Ho and Timothy~H Hsieh.
\newblock Efficient variational simulation of non-trivial quantum states.
\newblock \emph{SciPost Phys}, 6:\penalty0 029, 2019.

\bibitem[Johansson et~al.(2012)Johansson, Nation, and Nori]{johansson2012qutip}
J~Robert Johansson, PD~Nation, and Franco Nori.
\newblock Qutip: An open-source python framework for the dynamics of open
  quantum systems.
\newblock \emph{Computer Physics Communications}, 183\penalty0 (8):\penalty0
  1760--1772, 2012.

\bibitem[Johansson et~al.(2013)Johansson, Nation, and Nori]{johansson2013qutip}
J~Robert Johansson, Paul~D Nation, and Franco Nori.
\newblock Qutip 2: A python framework for the dynamics of open quantum systems.
\newblock \emph{Computer Physics Communications}, 184\penalty0 (4):\penalty0
  1234--1240, 2013.

\bibitem[Kakade(2002)]{kakade2002natural}
Sham~M Kakade.
\newblock A natural policy gradient.
\newblock In \emph{Advances in neural information processing systems}, pages
  1531--1538, 2002.

\bibitem[Kingma and Ba(2014)]{kingma2014adam}
Diederik~P Kingma and Jimmy Ba.
\newblock Adam: A method for stochastic optimization.
\newblock \emph{arXiv preprint arXiv:1412.6980}, 2014.

\bibitem[Kingma et~al.(2016)Kingma, Salimans, Jozefowicz, Chen, Sutskever, and
  Welling]{kingma2016improved}
Durk~P Kingma, Tim Salimans, Rafal Jozefowicz, Xi~Chen, Ilya Sutskever, and Max
  Welling.
\newblock Improved variational inference with inverse autoregressive flow.
\newblock In \emph{Advances in neural information processing systems}, pages
  4743--4751, 2016.

\bibitem[Kosut et~al.(2013)Kosut, Grace, and Brif]{kosut2013robust}
Robert~L Kosut, Matthew~D Grace, and Constantin Brif.
\newblock Robust control of quantum gates via sequential convex programming.
\newblock \emph{Physical Review A}, 88\penalty0 (5):\penalty0 052326, 2013.

\bibitem[Lloyd(1995)]{lloyd1995almost}
Seth Lloyd.
\newblock Almost any quantum logic gate is universal.
\newblock \emph{Physical Review Letters}, 75\penalty0 (2):\penalty0 346, 1995.

\bibitem[Lloyd(2018)]{lloyd2018quantum}
Seth Lloyd.
\newblock Quantum approximate optimization is computationally universal.
\newblock \emph{arXiv preprint arXiv:1812.11075}, 2018.

\bibitem[Matsumine et~al.(2019)Matsumine, Koike-Akino, and
  Wang]{matsumine2019channel}
Toshiki Matsumine, Toshiaki Koike-Akino, and Ye~Wang.
\newblock Channel decoding with quantum approximate optimization algorithm.
\newblock In \emph{2019 IEEE International Symposium on Information Theory
  (ISIT)}, pages 2574--2578. IEEE, 2019.

\bibitem[McClean et~al.(2018)McClean, Boixo, Smelyanskiy, Babbush, and
  Neven]{mcclean2018barren}
Jarrod~R McClean, Sergio Boixo, Vadim~N Smelyanskiy, Ryan Babbush, and Hartmut
  Neven.
\newblock Barren plateaus in quantum neural network training landscapes.
\newblock \emph{Nature communications}, 9\penalty0 (1):\penalty0 1--6, 2018.

\bibitem[Morales et~al.(2019)Morales, Biamonte, and
  Zimbor{\'a}s]{morales2019universality}
Mauro~ES Morales, Jacob Biamonte, and Zolt{\'a}n Zimbor{\'a}s.
\newblock On the universality of the quantum approximate optimization
  algorithm.
\newblock \emph{arXiv preprint arXiv:1909.03123}, 2019.

\bibitem[Nakanishi et~al.(2019)Nakanishi, Fujii, and
  Todo]{nakanishi2019sequential}
Ken~M Nakanishi, Keisuke Fujii, and Synge Todo.
\newblock Sequential minimal optimization for quantum-classical hybrid
  algorithms.
\newblock \emph{arXiv preprint arXiv:1903.12166}, 2019.

\bibitem[Niu et~al.(2019{\natexlab{a}})Niu, Boixo, Smelyanskiy, and
  Neven]{niu2019universal}
Murphy~Yuezhen Niu, Sergio Boixo, Vadim~N Smelyanskiy, and Hartmut Neven.
\newblock Universal quantum control through deep reinforcement learning.
\newblock \emph{npj Quantum Information}, 5\penalty0 (1):\penalty0 33,
  2019{\natexlab{a}}.

\bibitem[Niu et~al.(2019{\natexlab{b}})Niu, Lu, and Chuang]{niu2019optimizing}
Murphy~Yuezhen Niu, Sirui Lu, and Isaac~L Chuang.
\newblock Optimizing qaoa: Success probability and runtime dependence on
  circuit depth.
\newblock \emph{arXiv preprint arXiv:1905.12134}, 2019{\natexlab{b}}.

\bibitem[Peruzzo et~al.(2014)Peruzzo, McClean, Shadbolt, Yung, Zhou, Love,
  Aspuru-Guzik, and O’brien]{peruzzo2014variational}
Alberto Peruzzo, Jarrod McClean, Peter Shadbolt, Man-Hong Yung, Xiao-Qi Zhou,
  Peter~J Love, Al{\'a}n Aspuru-Guzik, and Jeremy~L O’brien.
\newblock A variational eigenvalue solver on a photonic quantum processor.
\newblock \emph{Nature communications}, 5:\penalty0 4213, 2014.

\bibitem[Porotti et~al.(2019)Porotti, Tamascelli, Restelli, and
  Prati]{porotti2019coherent}
Riccardo Porotti, Dario Tamascelli, Marcello Restelli, and Enrico Prati.
\newblock Coherent transport of quantum states by deep reinforcement learning.
\newblock \emph{Communications Physics}, 2\penalty0 (1):\penalty0 61, 2019.

\bibitem[Powell(1964)]{powell1964efficient}
Michael~JD Powell.
\newblock An efficient method for finding the minimum of a function of several
  variables without calculating derivatives.
\newblock \emph{The computer journal}, 7\penalty0 (2):\penalty0 155--162, 1964.

\bibitem[Rapin and Teytaud(2018)]{nevergrad}
J.~Rapin and O.~Teytaud.
\newblock {Nevergrad - A gradient-free optimization platform}.
\newblock \url{https://GitHub.com/FacebookResearch/Nevergrad}, 2018.

\bibitem[Romero et~al.(2017)Romero, Olson, and Aspuru-Guzik]{romero2017quantum}
Jonathan Romero, Jonathan~P Olson, and Alan Aspuru-Guzik.
\newblock Quantum autoencoders for efficient compression of quantum data.
\newblock \emph{Quantum Science and Technology}, 2\penalty0 (4):\penalty0
  045001, 2017.

\bibitem[Sauvage and Mintert(2019)]{Sauvage2019}
Frederic Sauvage and Florian Mintert.
\newblock {Optimal quantum control with poor statistics}.
\newblock 2019.
\newblock URL \url{http://arxiv.org/abs/1909.01229}.

\bibitem[Sch{\"{a}}fer et~al.(2020)Sch{\"{a}}fer, Kloc, Bruder, and
  L{\"{o}}rch]{Schafer2020}
Frank Sch{\"{a}}fer, Michal Kloc, Christoph Bruder, and Niels L{\"{o}}rch.
\newblock {A differentiable programming method for quantum control}.
\newblock 2020.
\newblock URL \url{http://arxiv.org/abs/2002.08376}.

\bibitem[Schulman et~al.(2015)Schulman, Levine, Abbeel, Jordan, and
  Moritz]{schulman2015trust}
John Schulman, Sergey Levine, Pieter Abbeel, Michael Jordan, and Philipp
  Moritz.
\newblock Trust region policy optimization.
\newblock In \emph{International conference on machine learning}, pages
  1889--1897, 2015.

\bibitem[Schulman et~al.(2017)Schulman, Wolski, Dhariwal, Radford, and
  Klimov]{schulman2017proximal}
John Schulman, Filip Wolski, Prafulla Dhariwal, Alec Radford, and Oleg Klimov.
\newblock Proximal policy optimization algorithms.
\newblock \emph{arXiv preprint arXiv:1707.06347}, 2017.

\bibitem[Shi et~al.(2001)]{shi2001particle}
Yuhui Shi et~al.
\newblock Particle swarm optimization: developments, applications and
  resources.
\newblock In \emph{Proceedings of the 2001 congress on evolutionary computation
  (IEEE Cat. No. 01TH8546)}, volume~1, pages 81--86. IEEE, 2001.

\bibitem[S{\o}rdal and Bergli(2019)]{sordal2019deep}
Vegard~B S{\o}rdal and Joakim Bergli.
\newblock Deep reinforcement learning for robust quantum optimization.
\newblock \emph{arXiv preprint arXiv:1904.04712}, 2019.

\bibitem[Streif and Leib(2019)]{streif2019training}
Michael Streif and Martin Leib.
\newblock Training the quantum approximate optimization algorithm without
  access to a quantum processing unit.
\newblock \emph{arXiv preprint arXiv:1908.08862}, 2019.

\bibitem[Sutton and Barto(2018)]{sutton2018reinforcement}
Richard~S Sutton and Andrew~G Barto.
\newblock \emph{Reinforcement learning: {An} introduction}.
\newblock MIT press, 2018.

\bibitem[Tannor et~al.(1992)Tannor, Kazakov, and Orlov]{tannor1992control}
David~J Tannor, Vladimir Kazakov, and Vladimir Orlov.
\newblock Control of photochemical branching: Novel procedures for finding
  optimal pulses and global upper bounds.
\newblock In \emph{Time-dependent quantum molecular dynamics}, pages 347--360.
  Springer, 1992.

\bibitem[Torlai et~al.(2018)Torlai, Mazzola, Carrasquilla, Troyer, Melko, and
  Carleo]{torlai2018neural}
Giacomo Torlai, Guglielmo Mazzola, Juan Carrasquilla, Matthias Troyer, Roger
  Melko, and Giuseppe Carleo.
\newblock Neural-network quantum state tomography.
\newblock \emph{Nature Physics}, 14\penalty0 (5):\penalty0 447--450, 2018.

\bibitem[Verdon et~al.(2019)Verdon, Arrazola, Br{\'a}dler, and
  Killoran]{verdon2019quantum}
Guillaume Verdon, Juan~Miguel Arrazola, Kamil Br{\'a}dler, and Nathan Killoran.
\newblock A quantum approximate optimization algorithm for continuous problems.
\newblock \emph{arXiv preprint arXiv:1902.00409}, 2019.

\bibitem[Wauters et~al.(2020)Wauters, Panizon, Mbeng, and
  Santoro]{wauters2020reinforcement}
Matteo~M Wauters, Emanuele Panizon, Glen~B Mbeng, and Giuseppe~E Santoro.
\newblock Reinforcement learning assisted quantum optimization.
\newblock \emph{arXiv preprint arXiv:2004.12323}, 2020.

\bibitem[Weinberg and Bukov(2017)]{weinberg2017quspin}
Phillip Weinberg and Marin Bukov.
\newblock Quspin: a python package for dynamics and exact diagonalisation of
  quantum many body systems part i: spin chains.
\newblock \emph{SciPost Phys}, 2\penalty0 (1), 2017.

\bibitem[Weinberg and Bukov(2019)]{weinberg2019quspin}
Phillip Weinberg and Marin Bukov.
\newblock Quspin: a python package for dynamics and exact diagonalisation of
  quantum many body systems. part ii: bosons, fermions and higher spins.
\newblock \emph{SciPost Phys.}, 7\penalty0 (arXiv: 1804.06782):\penalty0 020,
  2019.

\bibitem[Wierstra et~al.(2008)Wierstra, Schaul, Peters, and
  Schmidhuber]{wierstra2008natural}
Daan Wierstra, Tom Schaul, Jan Peters, and Juergen Schmidhuber.
\newblock Natural evolution strategies.
\newblock In \emph{2008 IEEE Congress on Evolutionary Computation (IEEE World
  Congress on Computational Intelligence)}, pages 3381--3387. IEEE, 2008.

\bibitem[Williams(1992)]{williams1992simple}
Ronald~J Williams.
\newblock Simple {Statistical Gradient-Following Algorithms for Connectionist
  Reinforcement Learning}.
\newblock \emph{Machine learning}, 8\penalty0 (3-4):\penalty0 229--256, 1992.

\bibitem[Wu et~al.(2019)Wu, Ding, Dong, and Wang]{wu2019learning}
Re-Bing Wu, Haijin Ding, Daoyi Dong, and Xiaoting Wang.
\newblock Learning robust and high-precision quantum controls.
\newblock \emph{Physical Review A}, 99\penalty0 (4):\penalty0 042327, 2019.

\bibitem[Yang et~al.(2017)Yang, Rahmani, Shabani, Neven, and
  Chamon]{yang2017optimizing}
Zhi-Cheng Yang, Armin Rahmani, Alireza Shabani, Hartmut Neven, and Claudio
  Chamon.
\newblock Optimizing variational quantum algorithms using pontryagin’s
  minimum principle.
\newblock \emph{Physical Review X}, 7\penalty0 (2):\penalty0 021027, 2017.

\bibitem[Zhang et~al.(2019)Zhang, Wei, Asad, Yang, and
  Wang]{zhang2019reinforcement}
Xiao-Ming Zhang, Zezhu Wei, Raza Asad, Xu-Chen Yang, and Xin Wang.
\newblock When reinforcement learning stands out in quantum control? a
  comparative study on state preparation.
\newblock \emph{arXiv preprint arXiv:1902.02157}, 2019.

\bibitem[Zhao et~al.(2020)Zhao, Carleo, Stokes, and
  Veerapaneni]{zhao2020natural}
Tianchen Zhao, Giuseppe Carleo, James Stokes, and Shravan Veerapaneni.
\newblock Natural evolution strategies and quantum approximate optimization.
\newblock \emph{arXiv preprint arXiv:2005.04447}, 2020.

\end{thebibliography}

\appendix

\newpage

\section{Trajectories on the Bloch sphere}
In this appendix, we visualize the \agentname{} algorithm's final policies learned in the Fig.~\ref{fig:nonoise-sq}. 

\begin{figure}[H]
    \centering
    \includegraphics[scale=0.36]{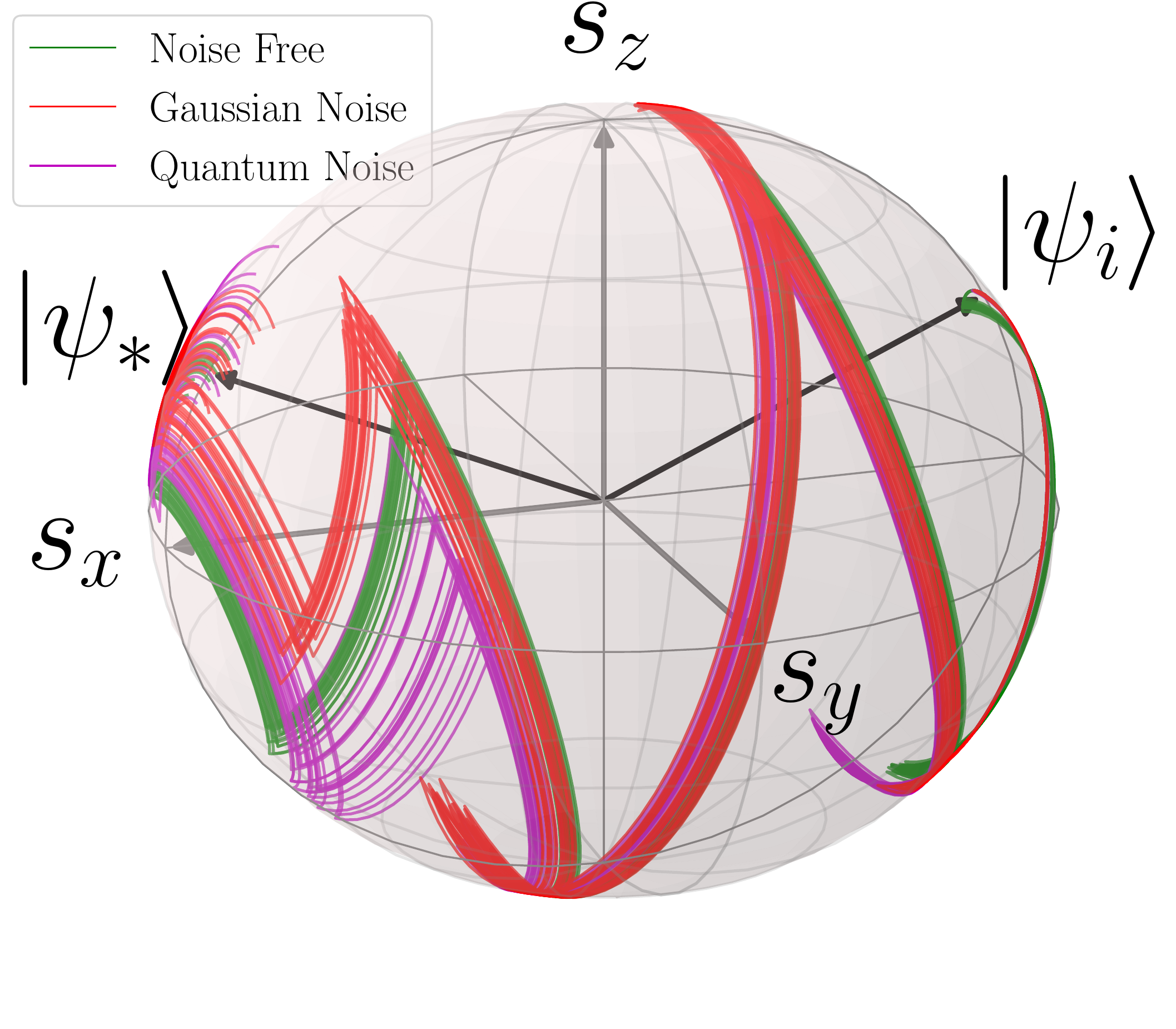}
    \vspace{-1em}
    \caption{\small Single-qubit testcase.  Trajectories of the protocols plotted on the Bloch sphere sampled from the learned policy. The three sets of curves correspond to the noise-free case (green), Gaussian noise (red), and quantum measurement noise (magenta). The simulation parameters are the same as in Fig.~\ref{fig:nonoise-sq}. 
    }
        \label{fig:trace-sq}
        \vspace{-0.5cm}
 \end{figure}
 
\newpage
\section{Learning curves: Gaussian noise}
\label{sec:app-gsfid}

In this appendix, we provide the learning curves for \agentname{} for various values of the standard deviation (i.e.~the noise level) of the Gaussian noise, and different number of qubits $N$.

\begin{figure}[H]
    \centerline{
    	 \subfigure[$\sigma = 0.01$, $(N, p) = (3, 15)$]{
        \includegraphics[scale=0.33]{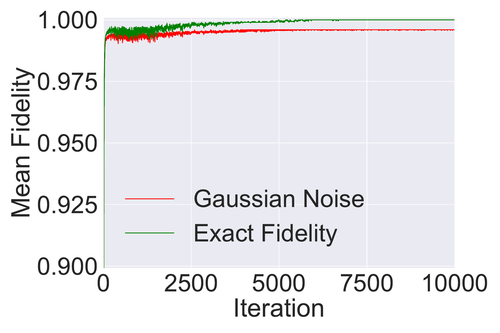}
        }
        \subfigure[$\sigma = 0.01$, $(N, p) = (4, 25)$]{
        \includegraphics[scale=0.33]{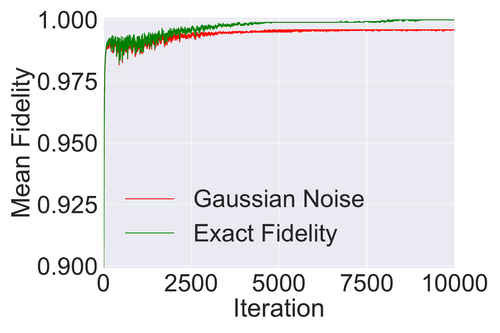}
	    }
	    \subfigure[$\sigma = 0.01$, $(N, p) = (5, 40)$]{
        \includegraphics[scale=0.33]{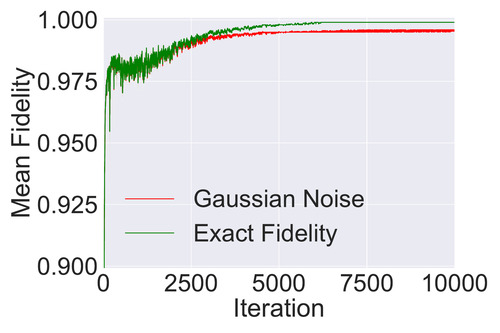}
	    }}
	    \centerline{
	   \subfigure[$\sigma = 0.05$, $(N, p) = (3, 15)$]{
        \includegraphics[scale=0.33]{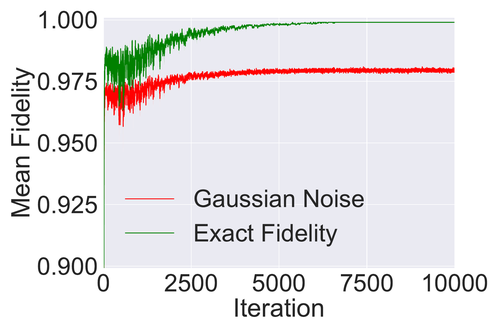}
        }
        \subfigure[$\sigma = 0.05$, $(N, p) = (4, 25)$]{
        \includegraphics[scale=0.33]{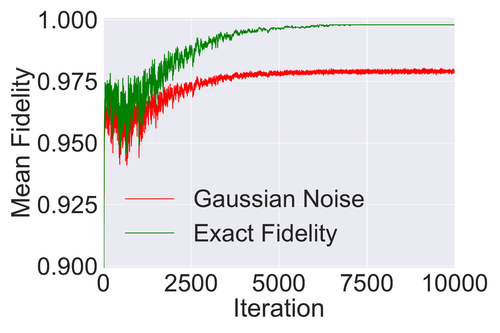}
	    }
	    \subfigure[$\sigma = 0.05$, $(N, p) = (5, 40)$]{
        \includegraphics[scale=0.33]{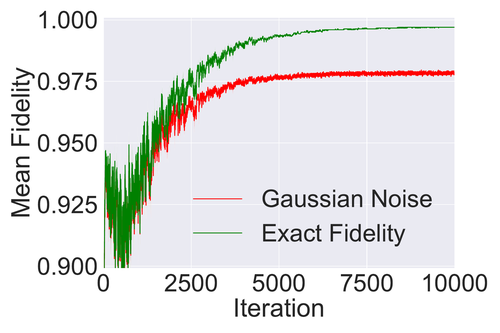}
	    }}
	    \centerline{
	    \subfigure[$\sigma = 0.1$, $(N, p) = (3, 15)$]{
        \includegraphics[scale=0.33]{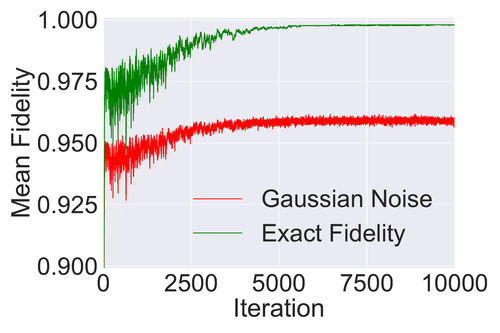}
        }
        \subfigure[$\sigma = 0.1$, $(N, p) = (4, 25)$]{
        \includegraphics[scale=0.33]{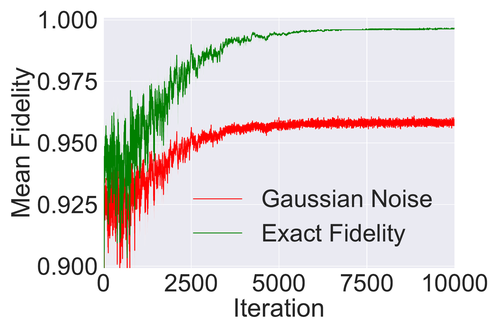}
	    }
	    \subfigure[$\sigma = 0.1$, $(N, p) = (5, 40)$]{
        \includegraphics[scale=0.33]{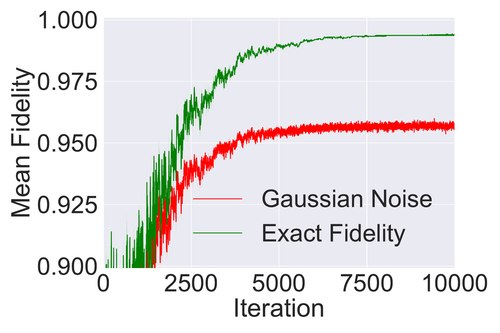}
	    }}
	\caption{\small Multi-quibit testcase I, Gaussain noise. Training curves (reward v.s. episode number) for various values of the Gaussian noise $\sigma$, and a few numbers of qubits $N$ and QAOA depth $p$. For the sake of comparison, we show the exact noise-free mean fidelity (green). The three rows: different Gaussian noise level. From top to bottom: $\sigma = 0.01$, $\sigma = 0.05$ and $\sigma = 0.1$. The three columns: different numbers of qubits $N$ and QAOA depth $p$. From left to right: $(N, p) = (3, 15)$, $(N, p) = (4, 25)$ and $(N, p) = (5, 40)$. The \agentname{} algorithm is trained with mini-batch sizes $M$ of 2048 for $10^4$ episodes. The means are initialized from a  truncated Gaussian distribution $\mathcal N(0.5, 0.1^2)$ and the stds are initialized to be 0.0024.}
	\label{fig:gsfid-plus}
	\vspace{-0.5cm}
 \end{figure}

\newpage
\section{Learning curves: quantum measurement noise}
\label{sec:app-qfid}

In this appendix, we provide the learning curves for \agentname{} for various values of the batch size used in the quantum measurement noise simulations, and different number of qubits $N$.

\begin{figure}[H]
     \centerline{
    	 \subfigure[$M = 128$, $(N, p) = (3, 15)$]{
        \includegraphics[scale=0.33]{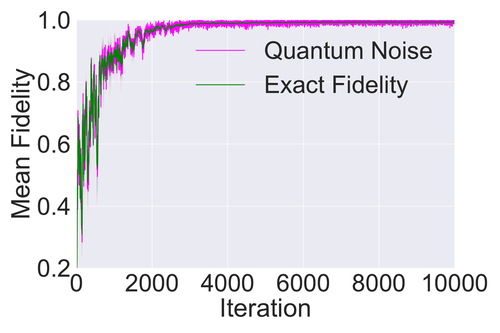}
        }
        \subfigure[$M = 128$, $(N, p) = (4, 25)$]{
        \includegraphics[scale=0.33]{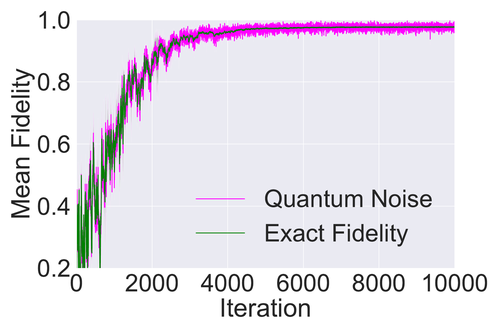}
	    }
	    \subfigure[$M = 128$, $(N, p) = (5, 40)$]{
        \includegraphics[scale=0.33]{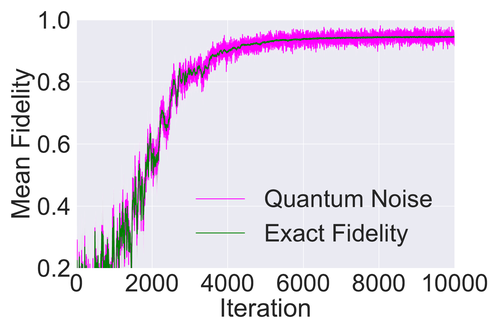}
	    }}
	    \centerline{
    	\subfigure[$M = 2048$, $(N, p) = (3, 15)$]{
        \includegraphics[scale=0.33]{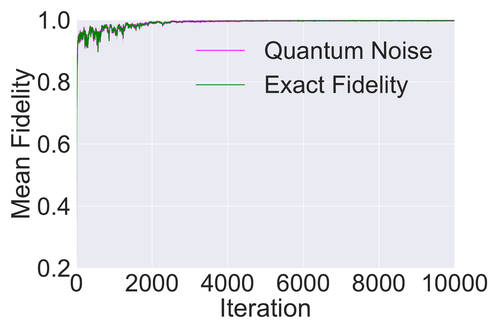}
        }
        \subfigure[$M = 2048$, $(N, p) = (4, 25)$]{
        \includegraphics[scale=0.33]{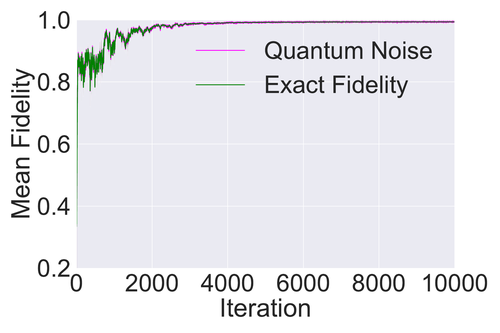}
	    }
	    \subfigure[$M = 2048$, $(N, p) = (5, 40)$]{
        \includegraphics[scale=0.33]{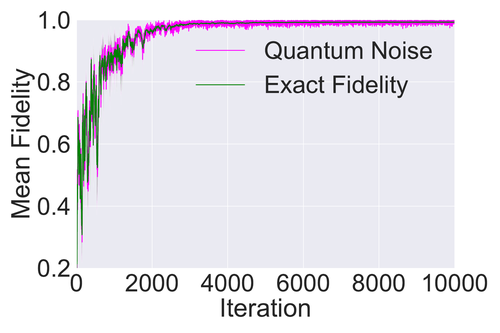}
	    }}
	\caption{\small Multi-quibit testcase I, quantum measurement noise. Training curves (reward v.s. episode number) for various values of the batch size $M$, and a few numbers of qubits $N$ and QAOA depth $p$. For the sake of comparison, we show the exact noise-free mean fidelity (green). The two rows: different mini-batch size $M$. From top to bottom: $M = 128$ and $M = 2048$. The three columns: different numbers of qubits $N$ and QAOA depth $p$. From left to right: $(N, p) = (3, 15)$, $(N, p) = (4, 25)$ and $(N, p) = (5, 40)$. The \agentname{} algorithm is trained for $10^4$ episodes. The means are initialized from a  truncated Gaussian distribution $\mathcal N(0.5, 0.1^2)$ and the stds are initialized to be 0.0024.
	}
	\label{fig:qfid-plus}
	\vspace{-0.5cm}
 \end{figure}
 
\section{Multivariate Gaussian policy with trainable covariance matrix}
\label{sec:app-multivariate}

In this appendix, we discuss a multivariate Gaussian policy. This is a natural generalization of the independent, decoupled Gaussians employed in the main text [Eq.~\ref{eqn:prob}]. The physical motivation for studying a policy which consists of correlated Gaussians is causality: the optimal action at some intermediate time step may well depend on the previous choices made during the episode. Such a dependence suggests the existence of correlations between the actions at different time steps. 

The easiest way to introduce correlations into the policy is to promote the policy ansatz from independent Gaussian distributions, to a single compound correlated Gaussian distribution. The correlations are modeled by the covariance matrix $\Sigma$. In the following, we shall refer to the uncorrelated policy as `diagonal', and the correlated policy -- as `non-diagonal' model. 

Let us define the correlated Gaussian policy $\pi_{\params}(\{\alpha_i, \beta_i\}_{i=1}^p)$ as
\begin{equation}
    \pi_{\params}(x)=\frac{1}{(2 \pi)^{p}| \det (\Sigma) |^{1 / 2}} \exp \left(-\frac{1}{2}(x-\mu)^{T} \Sigma^{-1}(x-\mu)\right).
    \label{eqn:cor-prob}
\end{equation}
Here, the sample $x = \{\alpha_i, \beta_i\}_{i=1}^p = \begin{pmatrix} \alpha_1 \\ \beta_1 \\ \vdots \\ \alpha_p \\ \beta_p \end{pmatrix}\in \mathbb R^{2p}$, and the trainable parameters are $\params = (\mu, \Sigma)$, where $\mu=\{ \mu_{\alpha_i}, \mu_{\beta_i}\}_{i=1}^p\in \mathbb R^{2p}$ and $\Sigma \in \mathbb R^{2p\times 2p}$.  The log likelihood for this policy reads 
\begin{equation}
    \log \pi_{\params}(x)=-\frac{1}{2} \log |\det(\Sigma)| -\frac{1}{2}(x-\mu)^{T} \Sigma^{-1}(x-\mu) + \text{const.}
    \label{eqn:cor-logprob}
\end{equation}

If $\Sigma=\operatorname{diag}(\sigma_{\alpha_{1}}^2, \sigma_{\beta_{1}}^2, \cdots , \sigma_{\alpha_{2p}}^2, \sigma_{\beta_{2p}}^2)$, this ansatz reduces to the uncorrelated policy used in the main text. The correlated policy is more expressive and we investigate its performance in the following experiments.

\subsection{Analytic expressions for the policy gradient}
The derivatives of the log likelihood with respect to the vector-valued $\mu$ and the matrix-valued $\Sigma$ can be computed as
\begin{eqnarray}
    \frac{\partial \log \pi_{\params}(x)}{\partial \mu}&=&\Sigma^{-1}(x-\mu),\\
    \frac{\partial \log \pi_{\params}(x)}{\partial \Sigma}&=&\frac 12 \left( - \Sigma^{-1} + \Sigma^{-1} (x-\mu)(x-\mu)^T\Sigma^{-1} \right).
    \label{eqn:cor-logprob-gradient}
\end{eqnarray}

Hence, the policy gradient in Eq.~(\ref{eqn:pg}) for the sample batch $B$ of size $M$ reads 
\begin{eqnarray}
     \nabla_{\mu}J(\params) &=& \frac{1}{M}\sum_{x_j \in B } \Sigma^{-1}(x_j-\mu) \cdot (F_j - \bar F), \\
           \nabla_{\Sigma}J(\params) &=&
           \frac{1}{2}\Sigma^{-1} \left( \frac{1}{M}\sum_{x_j \in B } (F_j - \bar F)\cdot (x_j-\mu)(x_j-\mu)^T  \right) \Sigma^{-1}, 
    \label{eqn:cor-pg}
\end{eqnarray}
where, as before, $\bar F$ is the mean of the fidelity reward $F_j$ over the sample $B$. Notice that the $\Sigma^{-1}$ term in Eq.~(\ref{eqn:cor-logprob-gradient}) does not depend on sample data, and thus vanishes as a consequence of using a baseline in the policy gradient algorithm [it is proportional to $\sum_{x_j \in B } (F_j - \bar F)=0$ ]. 

As a covariance matrix, $\Sigma$ should be both symmetric and positive-definite. Since any positive-definite matrix can be factorized as $\Sigma = AA^T$, we can use the matrix $A$ as a trainable (or learnable) variable. The corresponding gradient of the policy with respect to $A$ then reads

\begin{eqnarray}
           \nabla_{A}J(\params) &=&
          A^{-T} A^{-1}\left( \frac{1}{M}\sum_{x_j \in B } (F_j - \bar F)\cdot (x_j-\mu)(x_j-\mu)^T  \right) A^{-T}. 
           \label{eqn:cor-pg-A}
\end{eqnarray}

If we use the Cholesky factorization, then we may also restrict the matrix $A$ to be a lower-triangular matrix $L$, i.e.~$\Sigma = LL^T$. Then, Eq.~\ref{eqn:cor-pg-A} changes correspondingly to 

\begin{eqnarray}
           \nabla_{L}J(\params) &=& \operatorname{Tril}\left(
          L^{-T} L^{-1}\left( \frac{1}{M}\sum_{x_j \in B } (F_j - \bar F)\cdot (x_j-\mu)(x_j-\mu)^T  \right) L^{-T}\right).
    \label{eqn:cor-pg-L}
\end{eqnarray}

where $\operatorname{Tril}\left( \cdot \right)$ is the operation which outputs the lower triangular parts of any matrix. 

In the following discussion, we compare the learning behavior of PG-QAOA for the cases where $A$ is a diagonal matrix [i.e. the uncorrelated policy from the main text], with the two non-diagonal generalizations where (i) $A$ is an arbitrary matrix, and (ii) $A=L$ is lower-triangular.  

\subsection{Sampling a correlated Gaussian policy}
\label{sec:trans}

The sampling and probability density evaluation procedures for the multivariate Gaussian distribution are generalizations of those for uncorrelated normal distributions, by using a linear transformation as we now briefly explain. 

In order to sample the protocols according to the multivariate Gaussian policy, we first sample the i.i.d.~standard normal variables $z_i \sim \mathcal N(0, 1)$, or $z = (z_1, \dots, z_{2p})^T$. Then, we apply a linear transformation $f_\params(\cdot)$ to $z$:  $x = f_\params(z) = Az +\mu$.  In this way, the protocol samples $x$ are distributed according to the multivariate Gaussian $\mathcal N(\mu, \Sigma\!=\!AA^T)$, as required. Note that the policy parameters $\params$ we seek to learn, are actually the parameters of this transformation, i.e.~the matrix elements of $A$ and the vector elements of $\mu$.

Similarly, for the likelihood evaluation, the probability of $x$ can be calculated through the change-of-variable formula:

\begin{equation}
    \pi_\params(x)=\pi(z)\left|\operatorname{det}\left(\frac{\partial f(z)}{\partial z}\right)\right|^{-1} = \pi(z)\left|\operatorname{det}A \right|^{-1},
    \label{eqn:change-var-prob}
\end{equation}
where $\pi(z)=\frac{1}{(2 \pi)^{p}} \exp \left(-\frac{1}{2} z^{T} z\right)$ . 

The log likelihood then reads as
\begin{equation}
    \log \pi_\params(x) = \log \pi(z) - \log \left|\operatorname{det}A \right|.
    \label{eqn:change-var-logprob}
\end{equation}

Restricting the matrix $A=L$ to be a lower-triangular, one can readily build in casualty: for $A=L$ the $i$-th action (i.e.~step duration) of protocol $x_i$ only depends on the $z_1, z_2, \cdots, z_i$, i.e.~it only depends on the previous actions $x_1, \cdots, x_{i-1}$. Thus, if we view the policy as taking the actions in every RL episode sequentially, the action taken at a certain time step in the past will be able to influence all actions taken in the future.

Last, we note in passing that, as presented here, the multivariate Gaussian distribution serves as a specific case of the normalizing flow-based model~\citep{kingma2016improved, dinh2016density}. Generally, normalizing flows represent a larger class of distributions obtainable from the standard normal distribution. In that case, the transformation is parametrized via an invertible nonlinear neural network $f_\params(\cdot)$ instead of a linear transformation. The requirement of invertibility is because the Jacobian matrix has to be tractable or else we cannot uses Eq.~\ref{eqn:change-var-prob}.

\subsection{Numerical Experiments}

In this section, using the mutli-qubit systems, we investigate the performance of the multivariate Gaussian policy with learnable/trainable covariance matrix. 

The main observation is that the policy with covariance matrix can find protocols close to optimal when the number of qubits is small. We visualize the linear transformation matrix ($L$ and $A$) for the respective trained models. However, we find that training these correlated policies suffers from high variance when scaled to larger systems. This is due to the many trainable degrees of freedom contained in the covariance matrix. We also find that, in this situation, techniques like pretraining can provide a good initial point to stabilize the learning, yet the improvement over the uncorrelated policy is marginal.  

\subsubsection{Few-qubit systems}
\begin{figure}[h]
    \centerline{
    \includegraphics[width=0.49\textwidth]{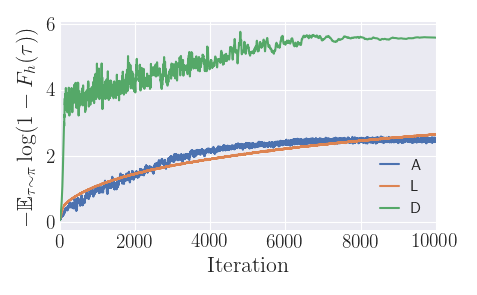}
    \includegraphics[width=0.49\textwidth]{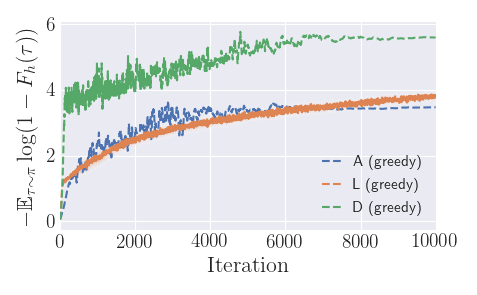}
    }
    \vspace{-1em}
    \caption{\small Noise-free Multi-qubit testcase I. Training (solid, left panel) and testing (dashed, right panel) curves for $N\!=\!3$ qubits with $p\!=\!15$. We compare (i) a diagonal Gaussian policy (D), (ii) a Gaussian policy with lower triangular  matrix (L), and (iii) a Gaussian policy with full  matrix (A). The metric on the $y$-axis is in log scale. During testing, only the means $\mu$  are used in a greedy evaluation. The learning rate for the diagonal (D) and full  matrix (A) is $5\cdot 10^{-4}$ with decay rate of $0.96$ every $50$ steps, and the learning rate for lower triangular (L) is $10^{-3}$ with decay rate of $0.99$ every $50$ steps. For hyperparameters cf.~caption of Fig.~\ref{fig:bo_comp}.
}
        \label{fig:N3-learn}
        \vspace{-0.3cm}
 \end{figure}

\begin{figure}[h]
    \centering
    \includegraphics[width=\textwidth]{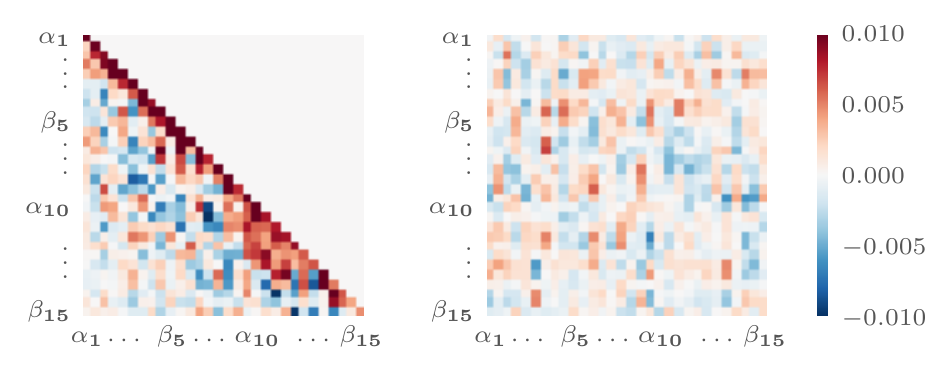}
    \vspace{-1em}
    \caption{\small Noise-free Multi-qubit testcase I. Visualization of the lower triangular matrix $L$ (left) and full matrix $A$ (right) for $N\!=\!3$ qubits with $p\!=\!15$. The data corresponds to the fully trained models from Fig.~\ref{fig:N3-learn}.}
        \label{fig:vis-cor}
        \vspace{-0.5cm}
 \end{figure}
 
 Fig.~\ref{fig:N3-learn} shows the training and testing curves for $N\!=\!3$ qubits, and QAOA circuit depth $p\!=\!15$. The fidelity from a  diagonal Gaussian policy (green) improves rapidly initially, and then converges gradually towards a high fidelity protocol. For the correlated Gaussian policies (blue and orange), the convergence of the fidelity is significantly slower. We checked that this behavior is not affected by reasonable choices of the learning rate and its decay schedule. In terms of the metric (i.e.~the negative log infidelity) in Fig.~\ref{fig:N3-learn}, we find there is more than $100\%$ gain in the training and $50\%$ gain in the testing when we use a diagonal Gaussian policy, as compared to the correlated policies. 
 
 After the training procedure finishes, we visualize the learned transformation matrices $L$ and $A$ shown in Fig.~\ref{fig:vis-cor}. As we mentioned in Sec.~\ref{sec:trans}, the actions given by these policies can be viewed as $x = Lz + \mu$ and $x = Az + \mu$, respectively.  
 Every time we sample independently  $z_i\sim \mathcal N(0, 1)$,  the policy produces a new action via $x_i = \sum_{j=1}^i L_{ij} z_j + \mu_i$ and $x_i = \sum_{j=1}^{2p} A_{ij} z_j + \mu_i$. The pixel color code (red denoting positive proportionality; blue -- negative proportionality) verifies our expectation that the correlated policies can learn interdependencies between specific time-durations $x_i$ in order to maximize the fidelity.  For the lower triangular matrix case, it also conveys causality, i.e. how the previous actions influence future ones. Interestingly, in Fig.~\ref{fig:vis-cor}, for the lower triangular policy (left), we observe that entries, close to the diagonal, dominate in magnitude -- as anticipated from the results on the diagonal policy discussed in the main text. When the distance in time becomes larger, their correlation also decays. This indicates that actions taken in the distant past rarely influence the future of an episode, in this problem.

\subsubsection{Behavior for larger systems}
 
 \begin{figure}[h]
    \centerline{
    \includegraphics[width=0.5\textwidth]{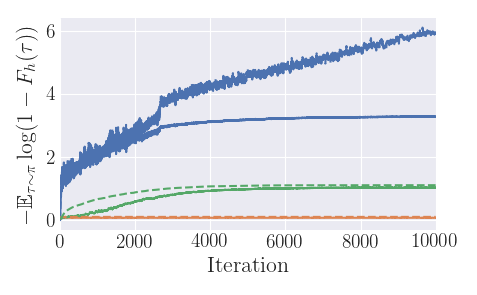}    
    \includegraphics[width=0.5\textwidth]{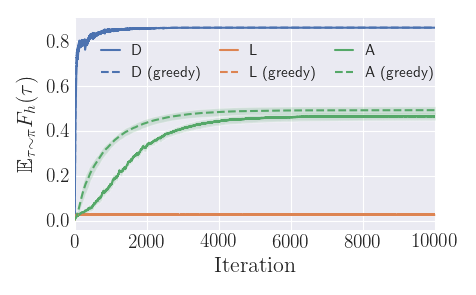}
    }
    \vspace{-1em}
    \caption{\small Noise-free Multi-qubit testcase I. Training (solid) and testing (dashed) curves for $N\!=\!6$ qubits (left) and $N\!=\!8$ qubits (right) with $p\!=\!60$. The metric is in log scale (left), and the normal scale (right). For the testing, only the means $\mu$  are used in a greedy evaluation.  Three policies are compared here, namely (i) diagonal Gaussian policy (D), (ii) Gaussian policy with lower triangular   matrix (L), and (iii) Gaussian policy with full  matrix (A). The learning rate for the diagonal/uncorrelated policy (D) is $10^{-2}$ and the learning rate for lower triangular (L) and full  matrix (A) is $10^{-3}$. Other hyperparameters are the same as in Fig.~\ref{fig:bo_comp}.}
        \label{fig:three-policy-comp}
        \vspace{-0.3cm}
 \end{figure}
 
A comparison among the three policies for $N=8$ qubits is shown in Fig.~\ref{fig:three-policy-comp}. The uncorrelated/diagonal Gaussian policy exhibits a fast convergence to a locally optimal protocol. 
The correlated Gaussian policy with lower-triangular matrix $L$ leads to early collapse of training and its performance remains the same as that of a random noisy policy. In contrast, the correlated Gaussian policy with full  matrix $A$ is able to find the right direction to improve the policy. 

This can be traced back to the major difference in the gradient of the policy for the two correlated policies -- the extra operator $\mathrm{Tril}(\cdot)$ removing the upper triangular parts in Eq.~\ref{eqn:cor-pg-L}. In the case of lower triangular matrix $L$, before applying this extra operation, the upper triangular parts of $L$  would still receive a non-zero gradient; the $\mathrm{Tril}(\cdot)$ operator sets these gradients to zero. 

Observe that, during training, the correlated Gaussian policy with full  matrix $A$ fails to achieve as high a fidelity as the diagonal/uncorrelated Gaussian policy. Since the number of parameters in the full covariance matrix policy case is the square of that in the diagonal/uncorrelated policy, we find that it is harder for gradient descent to control all these extra parameters in a way which substantially improves on the policy. Indeed, we observe that bad samples during the training can derail the learning process of the covariance matrix a lot. On the other hand, the diagonal/uncorrelated policy is more restricted, and hence prevents the distribution from being elongated in bad directions.

 \subsubsection{Pretraining}
 
\begin{figure}[h]
        \centerline{
    	 \subfigure[Gaussian Noise $\sigma = 0.00$, $(N, p) = (6, 60)$]{
        \includegraphics[scale=0.45]{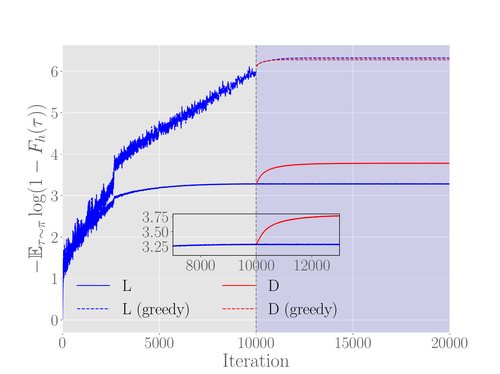}
        }
        \subfigure[Gaussian Noise $\sigma = 0.00$, $(N, p) = (8, 60)$]{
        \includegraphics[scale=0.45]{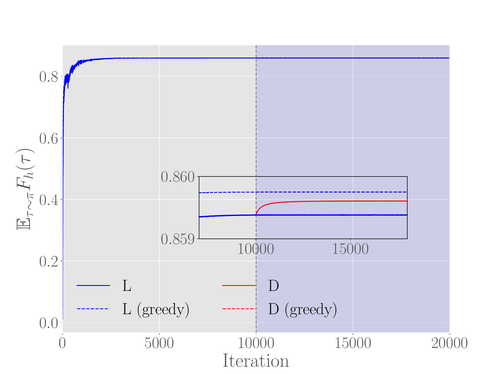}
	    }}
	    \centerline{
    	 \subfigure[Gaussian Noise $\sigma = 0.01$, $(N, p) = (6, 60)$]{
        \includegraphics[scale=0.45]{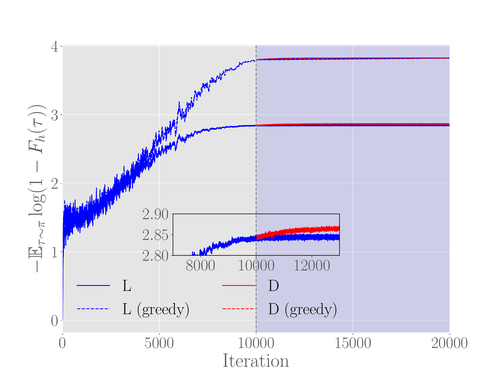}
        }
        \subfigure[Gaussian Noise $\sigma = 0.01$, $(N, p) = (8, 60)$]{
        \includegraphics[scale=0.45]{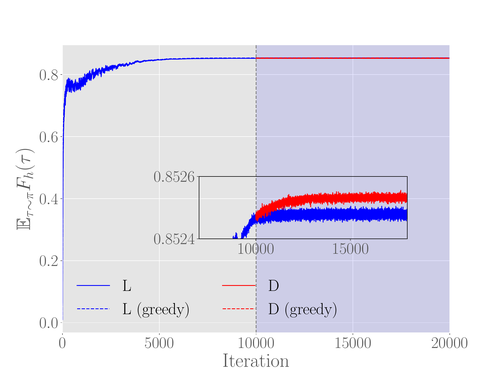}
	    }}
	    \centerline{
    	 \subfigure[Gaussian Noise $\sigma = 0.05$, $(N, p) = (6, 60)$]{
        \includegraphics[scale=0.45]{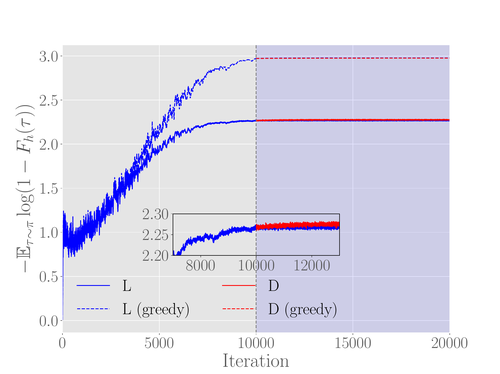}
        }
        \subfigure[Gaussian Noise $\sigma = 0.05$, $(N, p) = (8, 60)$]{
        \includegraphics[scale=0.45]{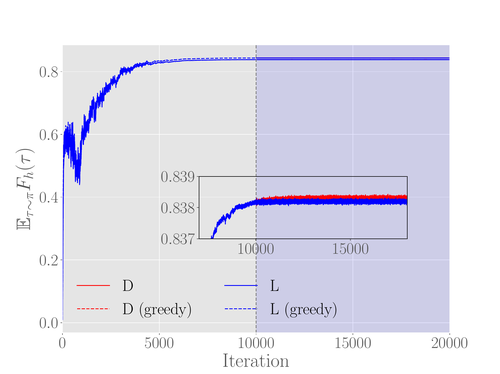}
	    }}
	\vspace{-0.3cm}
 \end{figure}
\begin{figure}[h]
        \ContinuedFloat
        \centerline{
    	 \subfigure[Gaussian Noise $\sigma = 0.1$, $(N, p) = (6, 60)$]{
        \includegraphics[scale=0.45]{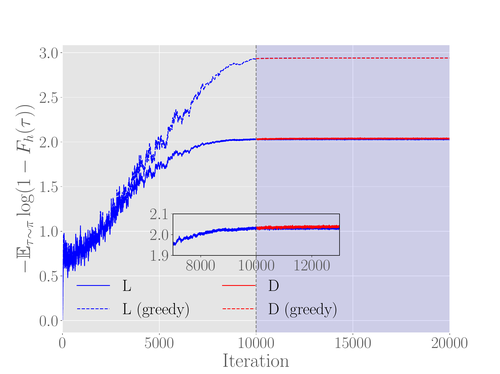}
        }
        \subfigure[Gaussian Noise $\sigma = 0.1$, $(N, p) = (8, 60)$]{
        \includegraphics[scale=0.45]{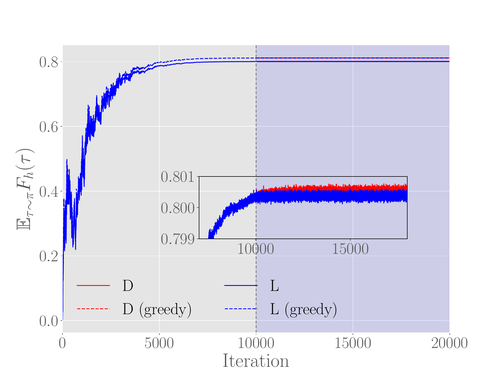}
	    }}
	    \centerline{
	   \subfigure[Quantum Noise ($Q$), $(N, p) = (6, 60)$]{
        \includegraphics[scale=0.45]{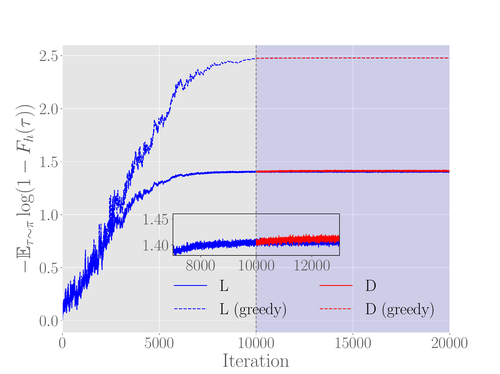}
        }
        \subfigure[Quantum Noise ($Q$), $(N, p) = (8, 60)$]{
        \includegraphics[scale=0.45]{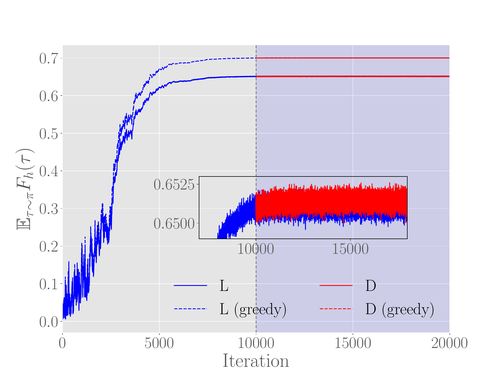}
	    }}
	\caption{\small Pretraining on Multi-qubit testcase I. Training and testing curves for $N\!=\!6$ qubits (left column) and $N\!=\!8$ qubits (right column) with $p\!=\!60$, and different fidelity noise level (cf. different standard deviations of the Gaussian noise and quantum measurement noise in the sub-captions). The $y$-axis metric is in log scale (left column, $L=6$), and in normal scale (right column, $L=8$). During testing, only the means $\mu$ are used in a greedy evaluation of a noise-free fidelity. During pretraining (i.e.~before $10^4$ episodes, light background shading), only the diagonal elements of the Gaussian policy are trained to get a good initialization. For the second part (dark background shading), two policies [uncorrelated/diagonal Gaussian policy (blue) and correlated policy with lower-triangular matrix $L$ (red)] are initialized with the learned parameters during pretraining. The hyperparameters for the pretraining are the same as in Fig.~\ref{fig:bo_comp}. The second parts are trained using SGD with learning rates of $10^{-2}$ and mini-batch size $M=2048$ for a total of $10^4$ episodes.}
	\label{fig:pretrain}
	\vspace{-0.3cm}
 \end{figure}

The problems with training correlated Gaussian policies with learnable off-diagonal covariance matrices, can be somewhat alleviated by using a pretraining procedure. Numerical experiments with pretraining based on the uncorrelated policy from the main text are shown in Fig.~\ref{fig:pretrain}. Their goal is to see if correlated policies can improve the quality of the solutions  found by the uncorrelated policy. 

As expected, pretraining provides a good initialization to stabilize the training of correlated Gaussian policies. However, we only found a marginal improvement of the fidelity using the correlated Gaussian policy compared to the uncorrelated one. We expect that the correlated Gaussian policy can achieve higher fidelity because 
the off-diagonal parts of the covariance matrix are also trainable parameters and the model is thus more expressive. 
The uncorrelated/diagonal policy used during pre-training already provides a good local minimum of the cost function landscape. The correlated Gaussian policy has more flexibility to change the shape of the sampling area which enables it to achieve higher training rewards over the diagonal policy. We find little to no benefit of using the correlated Gaussian policy in the problem of interest. One possible explanation for this behavior is the ability of the correlated Gaussian policy to deform its shape in order to fit the noise in the training batch; this in turn drags the policy mean vector $\mu$ away from its optimal value. Since only the means of the Gaussians are used for the greedy evaluation during the testing, correlated policies appear to perform worse.

Importantly, we find that the pretraining procedure is stable against various levels of Gaussian and quantum noise in the training data (Fig.~\ref{fig:pretrain}).

 \vspace{-0.3cm}

\end{document}